\documentclass[twocolumn,aps,prb]{revtex4-1}
\usepackage{amsmath}
\usepackage{amssymb}
\usepackage{braket}
\usepackage{graphicx}
\usepackage[table]{xcolor}
\begin{document}

\title{Electronic and optical properties of two-dimensional InSe\\ from a DFT-parameterized tight-binding model}
\author{S. J. Magorrian}
\author{V. Z\'{o}lyomi}
\author{V. I. Fal'ko} 
\affiliation{National Graphene Institute, University of Manchester, Booth St E, Manchester, M13 9PL, United Kingdom}
    \begin{abstract}
We present a tight-binding (TB) model and $\mathbf{k\cdot p}$ theory for electrons in monolayer and few-layer InSe. The model is constructed from a basis of all $s$ and $p$ valence orbitals on both indium and selenium atoms, with tight-binding parameters obtained from fitting to independently computed density functional theory (DFT) band structures for mono- and bilayer InSe. For the valence and conduction band edges of few-layer InSe, which appear to be in the vicinity of the $\Gamma$ point, we calculate the absorption coefficient for the principal optical transitions as a function of the number of layers, $N$. We find a strong dependence on $N$ of the principal optical transition energies, selection rules, and optical oscillation strengths, in agreement with recent observations \cite{Bandurin2016}. Also, we find that the conduction band electrons are relatively light ($m \propto 0.14-0.18 m_e$), in contrast to an almost flat, and slightly inverted, dispersion of valence band holes near the $\Gamma$-point, which is found for up to $N \propto 6$.
\end{abstract}
\maketitle

\section{Introduction}

Two-dimensional (2D) crystals are atomically thin films of van der Waals materials that are stable when exfoliated from the three-dimensional crystal due to the weak nature of the interaction holding the individual layers together\cite{Yoffe}. Examples of such materials include graphite\cite{Novoselov2005}, boron nitride\cite{Gorbachev2011}, and transition metal dichalcogenides\cite{Mak2010}, which have shown that properties of monolayer and bilayer crystals may strongly differ from the bulk properties of these layered compounds. Many of the transition metal dichalcogenides have been shown to possess optical properties that make them well suited for use in photodetectors and other optical or optoelectronic applications\cite{Splendiani2010,Korn2011,Wang2012,Xu2014,Jones2013,Gan2013,Wu2014,Sie2015,Wang2015,Liu2015}.

Another chalcogenide currently emerging as a high potential material for use on optical applications is the layered hexagonal metal chalcogenide InSe, atomically thin films of which are possible to fabricate\cite{Lei2014,Mudd2013,Mudd2014,Mudd2015,Tamalampudi2014,Balakrishnan2014}. While in its bulk form InSe\cite{Damon1954,Likforman1975,Williams1977,Manjon2004,Manjon2001,Pellicer-Porres1999,Goi1992,Kress-Rogers1982,Gorshunov1998,Dimitriev1995,sc10,Segura1983,Segura1997,McCanny1977,DeBlasi1983} is a direct gap semiconductor \cite{Camassel1978}, its electronic structure undergoes significant changes upon exfoliation to few-layer or monolayer thickness, with particularly interesting optical properties observed in recent experiments\cite{Bandurin2016, SciRep_2016}. Density functional theory (DFT) calculations for single layer crystals of  InSe\cite{Zolyomi2014,Rybkovskiy2014} predict a large increase in the band gap as compared to bulk crystals, with the valence band maximum slightly shifted from the $\Gamma$ point. Despite being a van der Waals layered material, bulk InSe has a light effective mass for electrons in the conduction and valence band across the layers. Therefore, it is expected that the band gap\cite{Zolyomi2014,Mudd2014,Sun} and related physical properties of few-layer InSe will exhibit a strong dependence on the number of layers.

In this work we develop a tight-binding (TB) model of atomically thin InSe, tracing the dependence of electronic and optical properties on the number of layers ($N$) in the film. We use density functional theory (DFT) to parametrize the model and apply a scissor correction to compensate for the underestimation of the band gap. Indeed, we find that as compared to the majority of other layered materials with van der Waals coupling between consecutive layers, which have the out-of-plane electron mass heavier than the in-plane mass, in InSe this relation is reversed leading to a strong $N$-dependence of the band gap. Also, the stacking of consecutive layers in few-layer $\gamma$-InSe similar to A-B-C stacking in graphite breaks the mirror-plane symmetry of monolayer InSe, which should be expected to affect optical selection rules and SO coupling in few-layer InSe. 

We use the TB model developed here to predict the band structure of few-layer InSe, and we develop a $\mathbf{k\cdot p}$ model to predict the optical properties, with the matrix elements of the momentum operator obtained from the TB model.  We provide estimates for the band edge optical absorption coefficient as a function of the number of layers. The paper is structured as follows.

In Section \ref{structure} we discuss the crystal structure of InSe. In Section \ref{the_model} we present the model of monolayer InSe and in Section \ref{fewlayer_crystal_struct} we expand it to bilayer InSe. In Section \ref{Few_layer_TBSC} we apply the model to few-layer InSe. In Section \ref{kp_section} we present a 4-band $\mathbf{k} \cdot \mathbf{p}$ theory model, and we calculate the momentum matrix elements and the band edge optical absorption in few-layer InSe, which enables us to interpret the recent experimental results in Ref.\onlinecite{Bandurin2016}. The $\mathbf{k\cdot p}$ theory in section \ref{kp_section} also describes spin-orbit coupling terms in mono- and few-layer InSe.

\section{Crystal Structure and symmetry}
\label{structure}

\begin{figure*}
 \centering
   \includegraphics[width=0.6\textwidth]{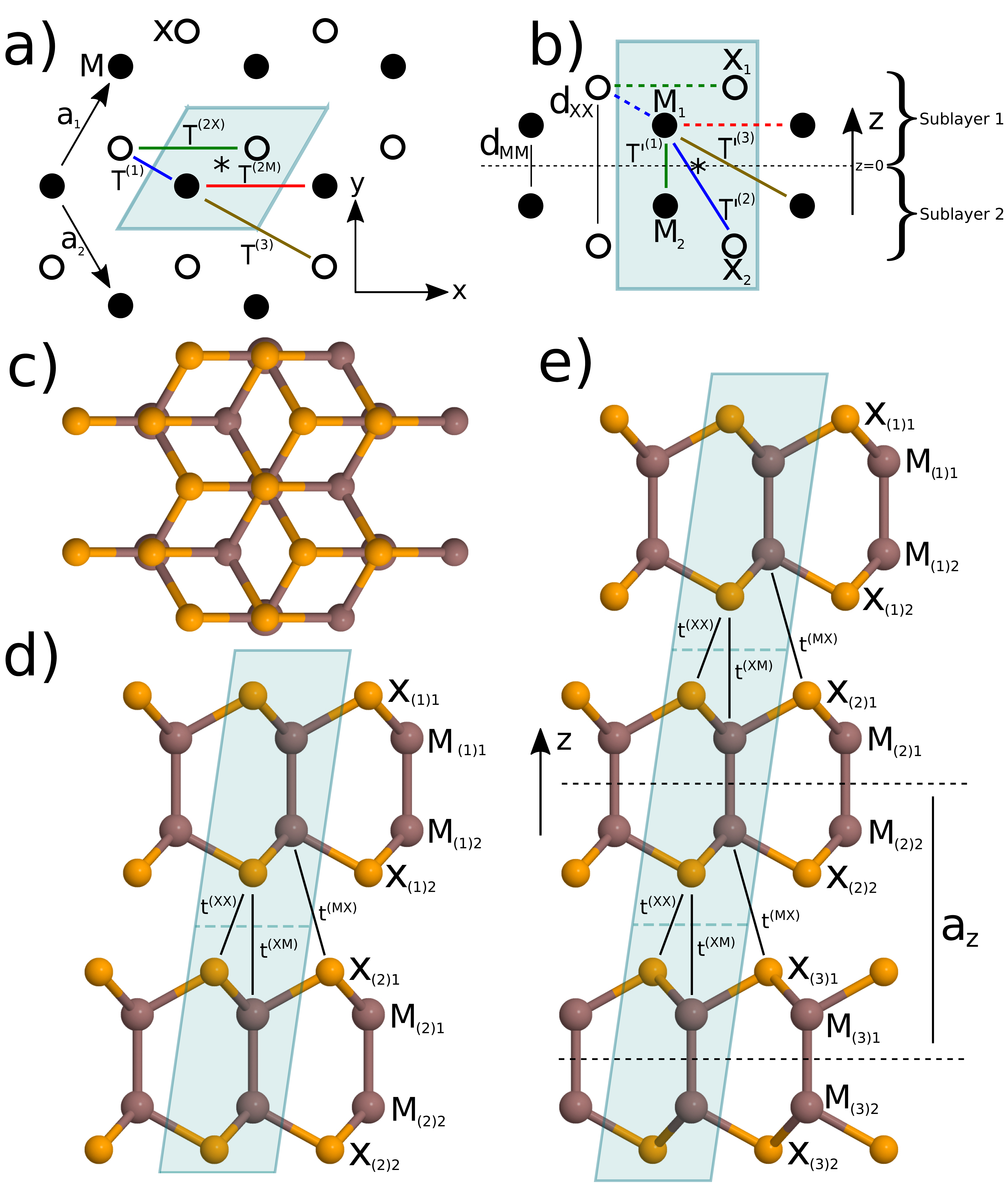}
 \caption{(Color online) Schematic of InSe illustrating a top view in the $xy$ plane (a) and a side view in the $xz$ plane (b) of the monolayer structure, indicating inequivalent hoppings included in the tight binding model, top (c) and side (d) views of the $\gamma$-stacked bilayer, and a side view of the trilayer crystal structure (e) with included interlayer TB hoppings indicated. Shaded region is the unit cell, with * indicating the chosen position of the unit cell origin. The In atoms are marked as M and Se atoms as X in the figure. The lattice parameters of the monolayer crystal according to the local density approximation\cite{Zolyomi2014} are $a=3.953$~\AA, $d_{MM}=2.741$~\AA, and $d_{XX}=5.298$~\AA.
 }
\label{fig_structure}
\end{figure*}

The crystalline structure of monolayer InSe considered in this study takes the form of hexagonal III-VI chalcogenides in M$_2$X$_2$ stoichiometry, where M is a metal atom of group III and X is a chalcogen atom of group VI. The structure is illustrated in Fig. \ref{fig_structure}a-c. A unit cell of the monolayer consists of four ions - one metal and one chalcogen in each of two sublayers.

The monolayer crystal has point-group symmetry $D_{3h}=C_{3v} \otimes \sigma_h$ (see Fig. \ref{Energy_modelDFT_a}) which includes $z\rightarrow -z$ mirror symmetry ($\mathcal{M}_1$, or $\sigma_h$ reflection). This symmetry operation effectively swaps the sublayers.  From a top-down view, the crystal exhibits a honeycomb structure in the $xy$ plane, where $A$ sites are occupied by metal ions and $B$ sites by chalcogen ions, possessing rotational symmetry centered at each atomic position ($\mathcal{R}_3$, or $C_3$ rotation) and mirror symmetry ($\mathcal{M}_2$, or $\sigma_v$ reflection) in the $yz$ plane (and equivalent planes generated by $\mathcal{R}_3$).
The Bravais lattice is given by
\begin{align}
\begin{split}
\mathbf{a_{1,2}}&=\frac{a}{2}\hat{\mathbf{x}}\pm\frac{\sqrt{3}a}{2}\hat{\mathbf{y}},\\
\mathbf{R}_i&=l_{1i}\mathbf{a_1}+l_{2i}\mathbf{a_2},
\end{split}
\end{align}

\noindent where $l_{1i}$ and $l_{2i}$ are integers, and the full crystal structure is given by

\begin{align}
\begin{split}
\mathbf{R}_{M1i}&=\mathbf{R}_{i}-\frac{a}{4}\left[\hat{\mathbf{x}}+\frac{\hat{\mathbf{y}}}{\sqrt{3}}\right]+\frac{d_{MM}}{2}\hat{\mathbf{z}},\\
\mathbf{R}_{M2i}&=\mathbf{R}_{i}-\frac{a}{4}\left[\hat{\mathbf{x}}+\frac{\hat{\mathbf{y}}}{\sqrt{3}}\right]-\frac{d_{MM}}{2}\hat{\mathbf{z}},\\
\mathbf{R}_{X1i}&=\mathbf{R}_{i}+\frac{a}{4}\left[\hat{\mathbf{x}}+\frac{\hat{\mathbf{y}}}{\sqrt{3}}\right]+\frac{d_{XX}}{2}\hat{\mathbf{z}},\\
\mathbf{R}_{X2i}&=\mathbf{R}_{i}+\frac{a}{4}\left[\hat{\mathbf{x}}+\frac{\hat{\mathbf{y}}}{\sqrt{3}}\right]-\frac{d_{XX}}{2}\hat{\mathbf{z}},
\end{split}
\end{align}

\noindent where $M_{1(2)i}/X_{1(2)i}$ is a In/Se atom in the top (bottom) sublayer. The structure of $\gamma$-InSe is shown in Fig. \ref{fig_structure}e. The monolayers are stacked such that chalcogen atoms in the top layer are directly above the metal atoms in the layer below, while the chalcogen atoms in the bottom layer are not directly below the metal atoms in the layer above. The vector between a chalcogen atom and the metal atom directly below it is

\begin{equation}
-\left[a_z-\frac{d_{XX}+d_{MM}}{2}\right]\mathbf{\hat{z}}
\end{equation}

\noindent while the vectors between a chalcogen atom and the nearest chalcogen atoms in the layer below are

\begin{equation}
-\left[a_z-\frac{d_{XX}+d_{MM}}{2}\right]\mathbf{\hat{z}}+\mathbf{r}_i
\end{equation}

\noindent where $\mathbf{r}_i$ ($i$=1,2,3) are the vectors between nearest-neighboring M-X pairs in the top sublayer of a monolayer. $a_z=$8.32~\AA~ is the distance along $z$ between the central $xy$ plane of each layer. The structure parameters for monolayer InSe are given in the caption of Fig. \ref{fig_structure}.

It is important to note here that, due to the stacking, the point symmetry of the material is reduced from that of the monolayer. Bulk and few-layer $\gamma$-InSe exhibit only $C_{3v}$ symmetry while in the monolayer we have $D_{3h}$ symmetry. The main difference between the two cases is that the $\mathcal{M}_1$ symmetry of the monolayer is broken by the stacking when we have more than one layer; this has important consequences for the optical matrix element, which is discussed below. In the bulk adjacent monolayers are related by $3_1$ and $3_2$ screw axes along $z$. The space group symmetry for the bulk crystal is $R3m$.

\section{TB model for monolayer InSe}
\label{the_model}

\subsection{Hamiltonian}
To describe InSe in a TB model, we construct our basis from the $s$ and $p$ orbitals of In (group III) and Se (group VI) atoms, and consider all possible hoppings between these orbitals up to second-nearest neighbor interactions. The Hamiltonian takes the form

\begin{equation}
H=\sum_{f}\left(H_{0f}+H_{ff}+H_{ff'}\right).
\end{equation}

\noindent where the sum over $f=1,2$ runs over the sublayers in the model, and $f'=2(1)$ when $f=1(2)$. Here, $H_{0f}$ contains terms arising from the on-site energies of the orbitals, while $H_{ff}$ and $H_{ff'}$ describe the hopping interactions within and between the sublayers, respectively, detailed below. Motivated by the dominant orbital contributions in DFT data for bands with energies near the Fermi level, we start from an atomic orbital basis including $s$ and $p$ orbitals in the valence shells of $M$ and $X$ atoms. $H_{0f}$ takes the form

\begin{align}
\begin{split}
H_{0f}=\sum_{i}&\left[\varepsilon_{Ms}m_{fis}^{\dagger}m_{fis}+\sum_{\alpha}\varepsilon_{p\alpha}m_{fip\alpha}^{\dagger}m_{fip\alpha}\right.\\
&\left.+\varepsilon_{Xs}x_{fis}^{\dagger}x_{fis}+\sum_{\alpha}\varepsilon_{p\alpha}x_{fip\alpha}^{\dagger}x_{fip\alpha}\right],
\end{split}
\end{align}

\noindent where the sum in $i$ goes over all unit cells in the crystal, while $\alpha=x,y,z$. Parameters $\varepsilon_{Ms}$ and $\varepsilon_{Xs}$ are on-site energies for the $s$ orbitals of metal and chalcogen ions respectively, while $\varepsilon_{Mp\alpha}$ and $\varepsilon_{Xp\alpha}$ are on-site energies for the relevant $p$ orbitals. $m_{fis}^{(\dagger)}$ is the annihilation (creation) operator for an electron in orbital $s$ on ion $M_f$ in unit cell $i$. $m_{fip\alpha}^{(\dagger)}$, is an annihilation (creation) operator for a $p_{\alpha}$ orbital.

$H_{ff}$ contains the hopping terms arising from intra-sublayer interactions, and is formed of the contributions 
\begin{equation}
H_{ff}=H_{ff}^{(1)}+H_{ff}^{(2M)}+H_{ff}^{(2X)}+H_{ff}^{(3)}
\end{equation}
where $H_{ff}^{(1)}$ includes nearest-neighbor hoppings for M-X pairs (labeled T$^{(1)}$ in Fig. \ref{fig_structure}), while $H_{ff}^{(2M)}$ and $H_{ff}^{(2X)}$ include hoppings for nearest pairs of like ions (M-M and X-X), (labeled T$^{(2M)}$ and T$^{(2X)}$), and $H_{ff}^{(3)}$ includes hoppings between next-nearest M-X pairs (labelled T$^{(3)}$). The contributions are
\begin{widetext}
\begin{align}
\begin{split}
H_{ff}^{(1)}=\sum_{<M_{fi},X_{fj}>}&\left[T^{(1)}_{ss}x_{fjs}^{\dagger}m_{fis}-T^{(1)}_{Ms-Xp}\sum_{\alpha}R^{M_{fi}X_{fj}}_{\alpha}x_{fjp\alpha}^{\dagger}m_{fis}+T^{(1)}_{Mp-Xs}\sum_{\alpha}R^{M_{fi}X_{fj}}_{\alpha}x_{fjs}^{\dagger}m_{fip\alpha}\right.\\
&\left.+\sum_{\alpha,\beta}\left(\left[\delta_{\alpha\beta}T^{(1)}_{\pi}-(T^{(1)}_{\pi}+T^{(1)}_{\sigma})R^{M_{fi}X_{fj}}_{\alpha}R^{M_{fi}X_{fj}}_{\beta}\right](x_{fjp\beta}^{\dagger}m_{fip\alpha})\right)\right]+h.c.
\end{split}
\end{align}
\begin{align}
\begin{split}
H_{ff}^{(2M)}=\sum_{<M_{fi},M_{fj}>}&\left[T^{(2M)}_{ss}m_{fjs}^{\dagger}m_{fis}-T^{(2M)}_{sp}\sum_{\alpha}R^{M_{fi}M_{fj}}_{\alpha}m_{fjp\alpha}^{\dagger}m_{fis}\right.\\
&\left.+\sum_{\alpha,\beta}\left(\left[\delta_{\alpha\beta}T^{(2M)}_{\pi}-(T^{(2M)}_{\pi}+T^{(2M)}_{\sigma})R^{M_{fi}M_{fj}}_{\alpha}R^{M_{fi}M_{fj}}_{\beta}\right](m_{fjp\beta}^{\dagger}m_{fip\alpha})\right)\right]+h.c.
\end{split}
\end{align}
\begin{align}
\begin{split}
H_{ff}^{(2X)}=\sum_{<X_{fi},X_{fj}>}&\left[T^{(2X)}_{ss}x_{fjs}^{\dagger}x_{fis}-T^{(2X)}_{sp}\sum_{\alpha}R^{X_{fi}X_{fj}}_{\alpha}x_{fjp\alpha}^{\dagger}x_{fis}\right.\\
&\left.+\sum_{\alpha,\beta}\left(\left[\delta_{\alpha\beta}T^{(2X)}_{\pi}-(T^{(2X)}_{\pi}+T^{(2X)}_{\sigma})R^{X_{fi}X_{fj}}_{\alpha}R^{X_{fi}X_{fj}}_{\beta}\right](x_{fjp\beta}^{\dagger}x_{fip\alpha})\right)\right]+h.c.
\end{split}
\end{align}
\begin{align}
\begin{split}
H_{ff}^{(3)}=\sum_{<M_{fi},X_{fj'}>}&\left[T^{(3)}_{ss}x_{fj's}^{\dagger}m_{fis}-T^{(3)}_{Ms-Xp}\sum_{\alpha}R^{M_{fi}X_{fj'}}_{\alpha}x_{fj'p\alpha}^{\dagger}m_{fis}+T^{(3)}_{Mp-Xs}\sum_{\alpha}R^{M_{fi}X_{fj'}}_{\alpha}x_{fj's}^{\dagger}m_{fip\alpha}\right.\\
&\left.+\sum_{\alpha,\beta}\left(\left[\delta_{\alpha\beta}T^{(3)}_{\pi}-(T^{(3)}_{\pi}+T^{(3)}_{\sigma})R^{M_{fi}X_{fj'}}_{\alpha}R^{M_{fi}X_{fj'}}_{\beta}\right](x_{fj'p\beta}^{\dagger}m_{fip\alpha})\right)\right]+h.c.
\end{split}
\end{align}
\end{widetext}

\noindent where the sum over $<M_{fi},X_{fj}>$ is over nearest-neighboring M-X pairs within a sublayer, and  $<M_{fi},M_{fj}>$,  $<X_{fi},X_{fj}>$,  and $<M_{fi},X_{fj'}>$ are over nearest M-M, X-X and next-nearest M-X pairs respectively. In considering the hoppings between the various $s$ and $p$ orbitals we have made the two-center approximation, as set out by Slater and Koster \cite{Slater1954}. $T^{(1)}_{ss}$ is the hopping integral for nearest-neighboring $s$ orbitals, $T^{(1)}_{Ms-Xp}$ and $T^{(1)}_{Mp-Xs}$ take into account $s-p$ hopping, while $T^{(1)}_{\pi}$ is the component of $p-p$ hopping where the $p$ orbitals are parallel to each other and perpendicular to the vector between the ions (hopping vector) and $T^{(1)}_{\sigma}$ is the hopping between the components of the $p$ orbitals lying along the hopping vector. $R^{M_{fi}X_{fj}}_{\alpha}$ takes account of the component of a $p$ orbital along the hopping vector, and thus has the form
\begin{equation}
R^{M_{fi}X_{fj}}_{\alpha}=\frac{\mathbf{R_{X_{fj}}}-\mathbf{R_{M_{fi}}}}{|\mathbf{R_{X_{fj}}}-\mathbf{R_{M_{fi}}}|}\cdot\hat{\boldsymbol{\alpha}}
\end{equation}
where $\hat{\boldsymbol{\alpha}}$ is a unit vector along $\alpha$.

The inter-sublayer hopping is written as
\begin{equation}
H_{ff'}=H_{ff'}^{(1)}+H_{ff'}^{(2)}+H_{ff'}^{(3)}
\end{equation}
where
\begin{widetext}
\begin{align}
\begin{split}
H_{ff'}^{(1)}=\sum_{i}&\left[T^{'(1)}_{ss}m_{f'is}^{\dagger}m_{fis}-T^{'(1)}_{sp}\sum_{\alpha}R^{M_{fi}M_{f'i}}_{\alpha}m_{fip\alpha}^{\dagger}m_{f'is}\right.\\
&\left.+\sum_{\alpha,\beta}\left(\left[\delta_{\alpha\beta}T^{'(1)}_{\pi}-(T^{'(1)}_{\pi}+T^{'(1)}_{\sigma})R^{M_{fi}M_{f'i}}_{\alpha}R^{M_{fi}M_{f'i}}_{\beta}\right](m_{f'ip\beta}^{\dagger}m_{fip\alpha})\right)\right],
\end{split}
\end{align}

\begin{align}
\begin{split}
H_{ff'}^{(2)}=\sum_{<M_{fi},X_{f'j}>}&\left[T^{'(2)}_{ss}x_{f'js}^{\dagger}m_{fis}-T^{'(2)}_{Ms-Xp}\sum_{\alpha}R^{M_{fi}X_{f'j}}_{\alpha}x_{2jp\alpha}^{\dagger}m_{1is}+T^{'(2)}_{Mp-Xs}\sum_{\alpha}R^{M_{fi}X_{f'j}}_{\alpha}x_{f'js}^{\dagger}m_{fip\alpha}\right.\\
&\left.+\sum_{\alpha,\beta}\left(\left[\delta_{\alpha\beta}T^{'(2)}_{\pi}-(T^{'(2)}_{\pi}+T^{'(2)}_{\sigma})R^{M_{fi}X_{f'j}}_{\alpha}R^{M_{fi}X_{f'j}}_{\beta}\right](x_{f'jp\beta}^{\dagger}m_{fip\alpha})\right)\right]+h.c.
\end{split}
\end{align}

\begin{align}
\begin{split}
H_{ff'}^{(3)}=\sum_{<M_{fi},M_{f'j}>}&\left[T^{'(3)}_{ss}m_{f'js}^{\dagger}m_{fis}-T^{'(3)}_{sp}\sum_{\alpha}R^{M_{fi}M_{f'j}}_{\alpha}m_{fjp\alpha}^{\dagger}m_{f'is}\right.\\
&\left.+\sum_{\alpha,\beta}\left(\left[\delta_{\alpha\beta}T^{'(3)}_{\pi}-(T^{'(3)}_{\pi}+T^{'(3)}_{\sigma})R^{M_{fi}M_{f'j}}_{\alpha}R^{M_{fi}M_{f'j}}_{\beta}\right](m_{f'jp\beta}^{\dagger}m_{fjp\alpha})\right)\right],
\end{split}
\end{align}

\end{widetext}

\subsection{DFT band structures of monolayer and bulk InSe and parametrization of monolayer TB model}
\label{model_vs_dft}
The DFT data to which we fit and compare the TB model of InSe in this section are obtained using the LDA exchange-correlation functional as set out in Ref. \onlinecite{Zolyomi2014} for In$_2$X$_2$ materials. In these calculations the VASP code\cite{VASP_PhysRevB.54.11169} is used to describe the materials in a plane-wave basis. The cutoff energy for the plane-wave basis is 600 eV and the vertical separation between repeated images of the monolayer is set to 20~\AA~ to ensure that any interactions between them would be negligible. The Brillouin zone is sampled by a $12 \times 12$ $\mathbf{k}$-point grid.

\begin{figure}
  \includegraphics[width=0.5\textwidth]{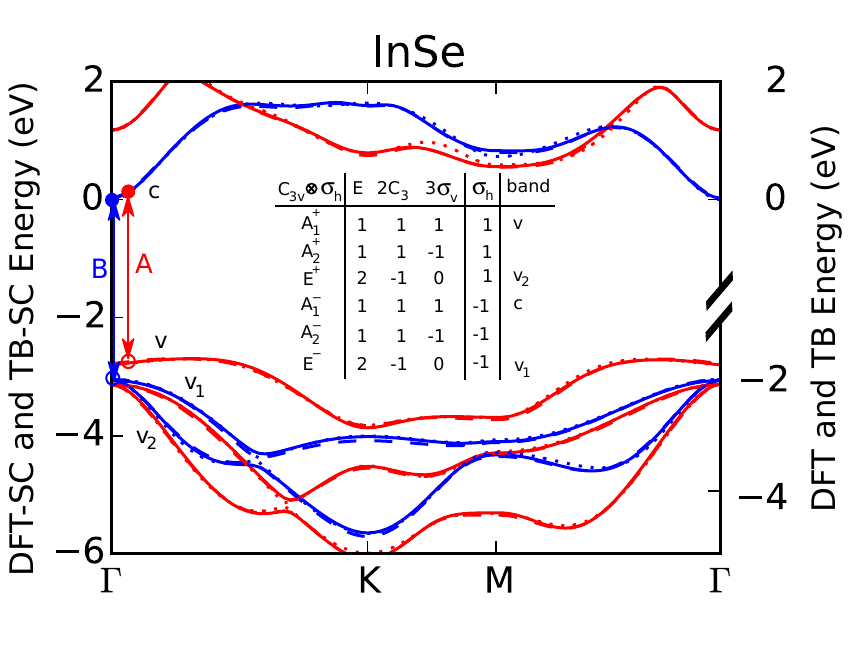}
  \caption{(Color online) Comparison between DFT and TB band structures of monolayer InSe. The TB band structure fitted to the scissor corrected DFT data (DFT-SC) is plotted with solid lines (TB-SC), the TB model fitted to the uncorrected DFT bands is plotted with dashed lines, and the DFT bands are plotted with dotted lines. Red lines are $z \rightarrow -z$ even bands, blue lines odd bands. Zero of energy set at the bottom of the conduction band. The left hand axis shows DFT energies after scissor correction of 0.99 eV (see text), the right hand axis the raw DFT energies. The lowest conduction band ($c$) and the highest three valence bands ($v$, $v_1$, $v_2$) are symmetry assigned according to their symmetry at the $\Gamma$-point, as determined by group theory. The character table for the irreducible representations of the point group $D_{3h}=C_{3v} \otimes \sigma_h$, with the irreps labeled by the names of the representations of point group $C_{3v}$ with a $+$ or $-$ superscript denoting the character of the $\sigma_h$ reflection, is shown in the inset. The TB fit to the scissor-corrected band structure is also shown (solid lines). Vertical lines marked $A$ and $B$ denote the principal optical transitions; of these, transition $A$ is forbidden by symmetry.}
\label{Energy_modelDFT_a}
\end{figure}

\begin{figure*}
  \includegraphics[width=0.90\textwidth]{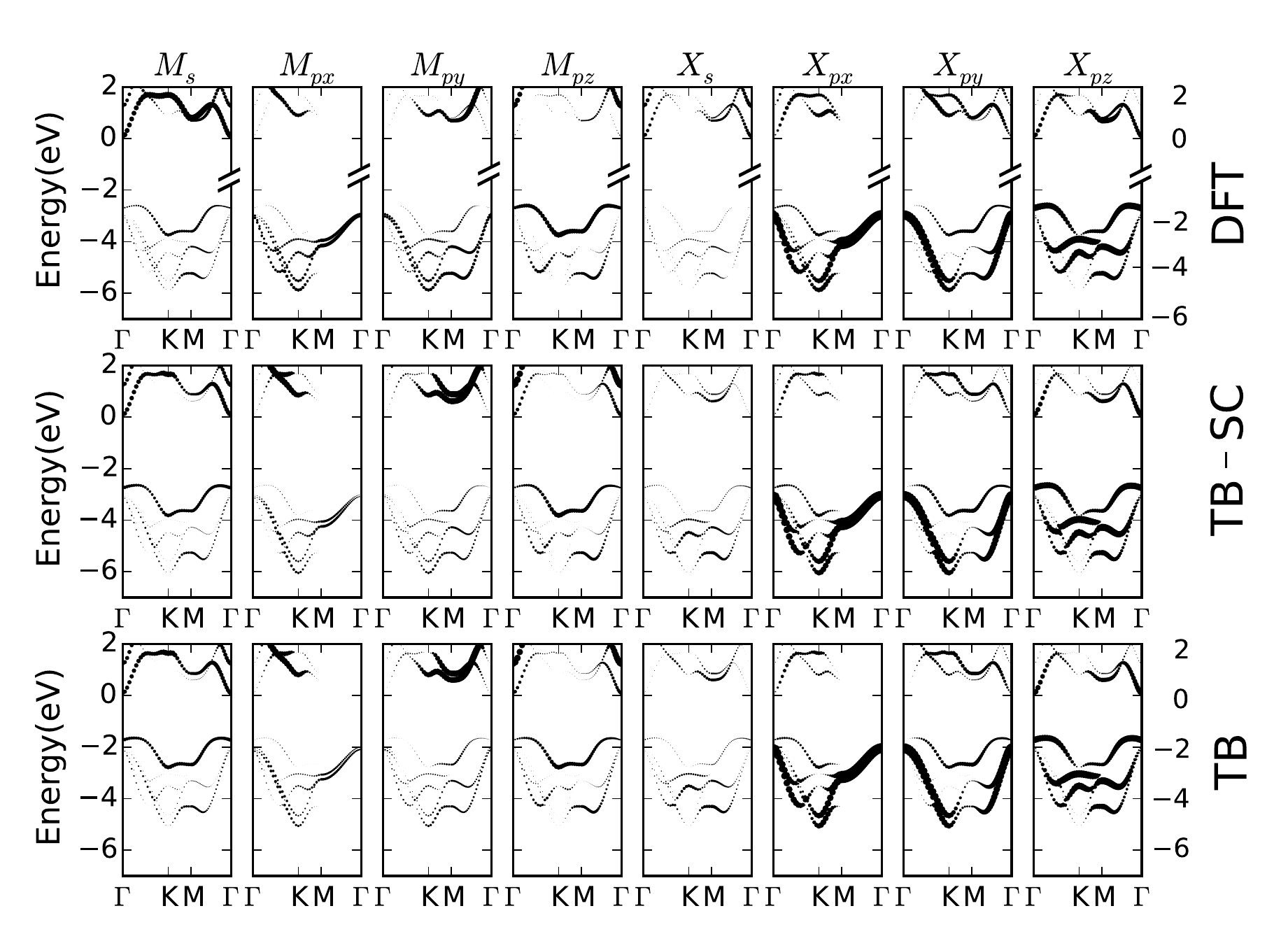}
  \caption{Orbital decomposition for fitted bands for InSe from both DFT data, and fitted model with (TB-SC) and without (TB) scissor correction. Marker size is proportional to normalized contribution.}
\label{Energy_modelDFT_b}
\end{figure*}

As semilocal density functional theory underestimates the band gap, we apply the ``scissor'' correction $\delta E_{g}$ to the DFT energy gaps, as employed before in the studies of other semiconductors \cite{sc3,sc4,sc5,sc6,sc7,sc8}, as follows. A calculation with the LDA returns the band gap for bulk InSe as 0.41 eV as compared to the bulk experimental value of 1.40 eV at low temperature \cite{sc9,sc10} (1.25 eV at room temperature \cite{Mudd2013}). Hence, we subtract $\delta E_{g}^{0K}\approx 0.99$~eV from the energies of all valence band states while keeping the conduction band energies unchanged for bulk, few-layer, and monolayer InSe. For optics, this scissor correction is equivalent to adding $\delta E_{g}^{0K}$ to the energies of all interband transitions (labeled $A$ and $B$ in Fig. \ref{Energy_modelDFT_a}), which we identify upon analyzing wave functions in the bands of monolayer and few-layer InSe (at room temperature, we use $\delta E_g^{300K}\approx 0.84$~eV).

In the following we fit the TB model to the scissor corrected DFT band structure, and call this scissor corrected tight-binding (TB-SC). We do this by applying a constrained least squares minimization procedure to the difference between the TB and the scissor corrected DFT band energies. While the procedure is in principle straightforward, in practice one must take care, in particular with the choice of bands to use for the fitting procedure. For comparison, we also perform a fit to the original DFT data to obtain a parametrization without scissor correction (TB).

On diagonalization the model yields 16 bands -- 8 even (symmetric) and 8 odd (antisymmetric) under $z\rightarrow -z$. As one progresses further in energy away from the conduction band edge and valence band edge the assumption that $s$ and $p$ orbital contributions dominate begins to break down, with significant $d$ orbital contributions at energies far away from the band edges. In addition, the DFT calculation is less accurate in the higher energy unoccupied bands. We therefore fit the model to 7 DFT bands - the 5 highest energy valence bands (3 even, 2 odd) and the 2 lowest energy conduction bands (1 even, 1 odd), using bands 3-6 of the 8 even model bands, and bands 3-5 of the odd model bands. As our primary purpose is a good quantitative fit to the valence and conduction band edges, we give extra weights to these points during the fitting procedure. The fit is carried out over a grid of 141 points in $\mathbf{k}$-space covering the irreducible portion of the Brillouin zone.

Table \ref{parameters} presents the parameters obtained in the fit for InSe with scissor correction taken into account (TB-SC) and without it (TB) for sake of completeness. Fig. \ref{Energy_modelDFT_a} shows the TB band structure (TB-SC) and the DFT data (DFT-SC) to which the fit was applied for InSe; the TB band structure without scissor correction (TB) is also plotted in comparison to the raw DFT data (DFT). The model gives a good reproduction of the DFT bands, both with and without scissor correction. Note that slight differences can be found between the shape of the fitted bands when comparing the fit to the raw DFT data and the scissor corrected bands, and the parameter sets differ accordingly.
\begin{table}
 \centering
\caption{Fitted parameters (eV) for the TB model of InSe based on DFT data with (TB-SC) and without (TB) scissor correction, as shown in Figure \ref{Energy_modelDFT_a}.}
\label{parameters}
\begin{tabular}{ccc}
\hline 
\hline 
 & TB-SC & TB\\  \hline
$\varepsilon_{M_{s}}$ & -7.174 & -7.595\\
$\varepsilon_{M_{px}}=\varepsilon_{M_{py}}$ & -2.302 & -3.027\\
$\varepsilon_{M_{pz}}$ & 1.248 & 0.903\\
$\varepsilon_{X_{s}}$ & -14.935 & -15.188\\
$\varepsilon_{X_{px}}=\varepsilon_{X_{py}}$ & -7.792 & -8.045\\
$\varepsilon_{X_{pz}}$ & -7.362 & -7.615\\ \hline
$T_{ss}^{(1)}$ & 0.168 & 0.331\\
$T_{Ms-Xp}^{(1)}$ & 2.873 & 2.599\\
$T_{Mp-Xs}^{(1)}$ &-2.144 & -2.263\\
$T_{\pi}^{(1)}$ & 1.041 & 0.977\\
$T_{\sigma}^{(1)}$ & 1.691 & 1.342\\ \hline

$T_{ss}^{(2M)}$ & -0.200 & -0.248\\
$T_{sp}^{(2M)}$ & -0.137 & -0.113\\
$T_{\pi}^{(2M)}$ & -0.433 & -0.561\\
$T_{\sigma}^{(2M)}$ & -1.034 & -1.130\\ \hline
$T_{ss}^{(2X)}$ & -1.345& -1.451\\
$T_{sp}^{(2X)}$ & -0.800 & -0.843\\
$T_{\pi}^{(2X)}$ & -0.148 & -0.110\\
$T_{\sigma}^{(2X)}$ & -0.554 & -0.613\\ \hline
$T_{ss}^{(3)}$ &0.821 & 0.793\\
$T_{Ms-Xp}^{(3)}$ & 0.156 & 0.179\\
$T_{Mp-Xs}^{(3)}$ &-0.294 & -0.323\\
$T_{\pi}^{(3)}$ & 0.003 & -0.015\\
$T_{\sigma}^{(3)}$ & -0.455 & -0.477\\ \hline
$T_{ss}'^{(1)}$ & -0.780 & -0.518\\
$T_{sp}'^{(1)}$ & -4.964 & -4.644\\
$T_{\pi}'^{(1)}$ & -0.681 & -0.769\\
$T_{\sigma}'^{(1)}$ & -4.028 & -4.052\\ \hline
$T_{ss}'^{(2)}$ & 0.574 & 0.472\\
$T_{Ms-Xp}'^{(2)}$ &-0.651& -0.544\\
$T_{Mp-Xs}'^{(2)}$ & -0.148& -0.138\\
$T_{\pi}'^{(2)}$ & 0.100 & 0.082\\
$T_{\sigma}'^{(2)}$ & 0.343  & 0.373\\ \hline
$T_{ss}'^{(3)}$ & -0.238& -0.187\\
$T_{sp}'^{(3)}$ & -0.048 & -0.065\\
$T_{\pi}'^{(3)}$ & -0.020 & -0.052\\
$T_{\sigma}'^{(3)}$ & -0.151 & -0.168\\ \hline
\hline
\end{tabular}
\end{table}

Alongside the energies predicted by our model Hamiltonian, it is useful to check the orbital decomposition, found in the normalized eigenvectors, against that given by the DFT results. We define $C_{n\mathbf{k}}(o)$ as the coefficient of the eigenfunction of band $n$, orbital $o$, at wave vector $\mathbf{k}$. Fig. \ref{Energy_modelDFT_b} shows the results of such a comparison for InSe, between the modulus square of the overlap integral between the DFT wave function and the spherical
harmonics centered on each atom, normalized against the total of $s$ and $p$ orbitals, and the equivalent $|C_{n\mathbf{k}}(o)|^2$  as calculated in the TB model. Larger markers indicate a more dominant contribution.
Table \ref{Tab_DFTorbital} gives the numerical contributions for the conduction band ($c$ ($z\rightarrow -z$ odd)), the valence band ($v$ ($z\rightarrow -z$ even)) and the next two (twice degenerate) bands just below the valence band at $\Gamma$.
We obtain a reasonable qualitative agreement between the model and DFT results.
\begin{table*}
\caption{The relative spherical harmonic character of the plane-wave wave function
(modulus square of the overlap integral between the DFT wave function and the spherical
harmonics centered on each atom) on the valence $s$ and $p$ orbitals of In and Se
atoms in monolayer InSe at the $\Gamma$-point, in the conduction band ($c$), the valence band ($v$),
and the two twice degenerate bands just below the valence band ($v_{1}$ and $v_{2}$), as labeled in Fig. \ref{Energy_modelDFT_a}. The equivalent contribution found in the scissor corrected TB model is given in square brackets. The $\Gamma$-point symmetry classification of the bands is noted in brackets.
Atoms are listed in order of increasing $z$ coordinate. Band energies (in eV) are provided relative
to the conduction band edge, where $E^{DFT}$ is the band energy from DFT and $E^{DFT-SC}=E^{DFT}-\delta E_{g}^{0K}$
is the value obtained after applying scissor correction.
The energies corresponding to $\hbar \omega _A$ and  $\hbar \omega _B$ are marked in bold.}
\label{Tab_DFTorbital}
\begin{center}
\begin{tabular}{lllll}

\hline
\hline
&$c$ $(A_1^{-})$&$v$ $(A_1^{+})$&$v_{1}$ $(E^{-})$&$v_{2}$ $(E^{+})$\\
\hline
$E^{DFT}$ (eV)& 0&  $-1.80$ &  $-2.05$ & $-2.13$ \\
\rowcolor{lightgray} $E^{DFT-SC}$ (eV)& &\textbf{-2.79}&\textbf{-3.04}&-3.12\\
Se1& 0.09[0.00]$s$  & 0.00[0.01]$s$  & 0.22[0.24]$p_{x(y)}$& 0.21[0.24]$p_{x(y)}$\\
   & 0.17[0.22]$p_z$& 0.35[0.36]$p_z$& & \\
In1& 0.23[0.16]$s$  & 0.03[0.10]$s$  & 0.03[0.01]$p_{x(y)}$& 0.04[0.01]$p_{x(y)}$\\
   & 0.01[0.12]$p_z$& 0.12[0.02]$p_z$& & \\
In2& 0.23[0.16]$s$  & 0.03[0.10]$s$  & 0.03[0.01]$p_{x(y)}$& 0.04[0.01]$p_{x(y)}$\\
   & 0.01[0.12]$p_z$& 0.12[0.02]$p_z$& & \\
Se2& 0.09[0.00]$s$  & 0.00[0.01]$s$  & 0.22[0.24]$p_{x(y)}$& 0.21[0.24]$p_{x(y)}$\\
   & 0.17[0.22]$p_z$& 0.35[0.36]$p_z$& & \\

\hline
\hline

\end{tabular}
\end{center}
\end{table*}
\subsection{Spin-orbit coupling}
\begin{figure}
\begin{center}
\includegraphics[width=0.5\textwidth]{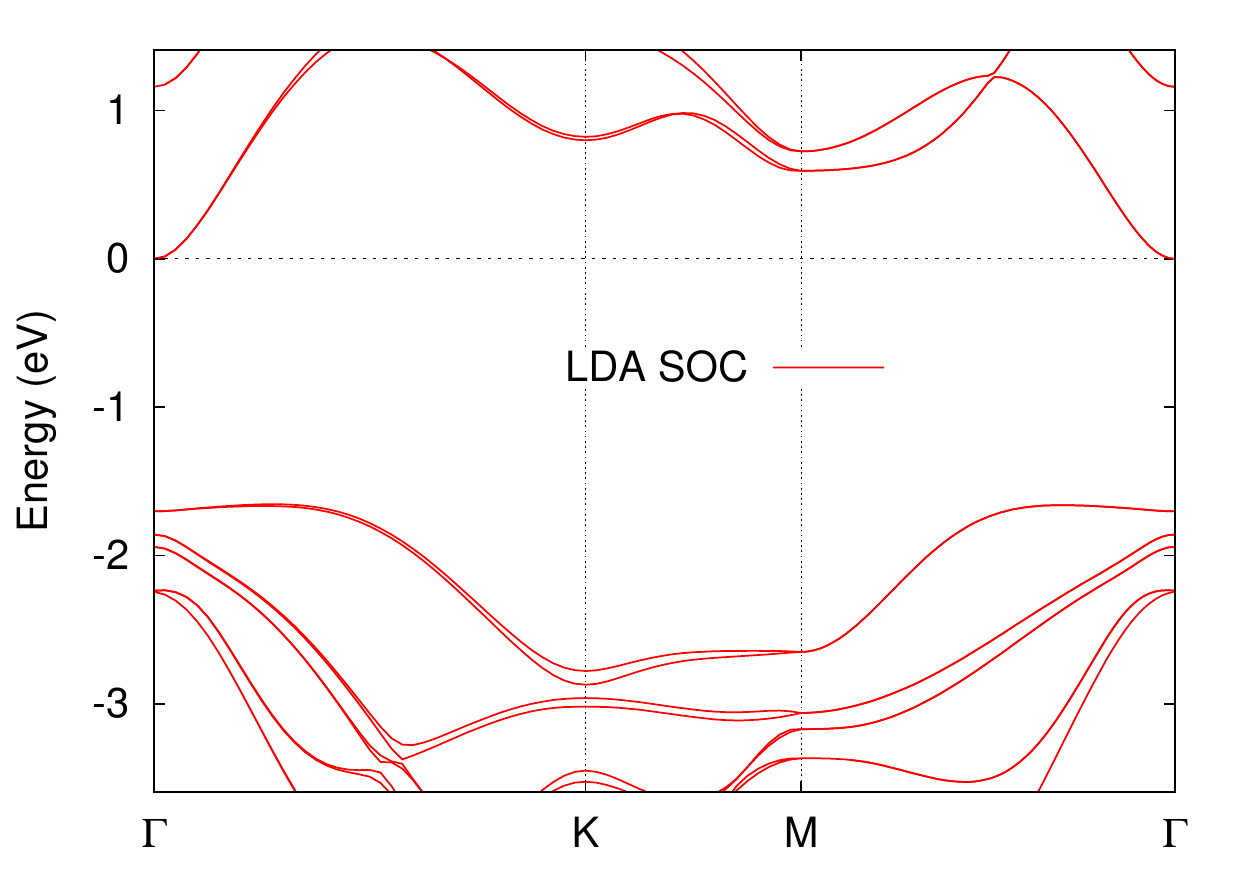}
\caption{LDA band structure of 1L-InSe with spin-orbit coupling taken into account. 
\label{FigS4}
}
\end{center}
\end{figure}
Fig. \ref{FigS4} shows the LDA band structure of the monolayer with spin-orbit coupling taken into account. The splitting is small, particularly so in the region of the $\Gamma$-point, we therefore neglect it in the TB model. In the $\mathbf{k\cdot p}$ theory (see section \ref{kp_section}), we include spin-orbit coupling to quantify how small it is near the $\Gamma$-point, and to show how it is expected to behave at a larger number of layers.

\section{Bilayer InSe: inter-layer hopping parameterized using DFT}
\label{fewlayer_crystal_struct}

We now extend the TB model to describe coupling between consecutive layers in $N$-layer InSe. For this, we consider a bilayer and include hops in the $z$-direction, XX, XM, and MX as depicted in Fig. \ref{fig_structure}d. The Hamiltonian can be written as 

\begin{equation}
H=H_{1}+H_{2}+H_{1,2}
\end{equation}

\noindent where $H_1$ and $H_2$ describe the individual monolayers comprising the bilayer structure, and $H_{1,2}$ describes the interaction between them and can be written as

\begin{equation}
H_{1,2}=H_{X_1,M_{2}}+H_{M_1,X_{2}}+H_{X_1,X_{2}}
\label{bilayer_hamiltonian}
\end{equation}

\noindent where each term corresponds to a category of hopping interactions as labeled in Fig. \ref{fig_structure}d. The vertical M-X contribution is

\begin{align}
\begin{split}
H_{X_1,M_{2}}=\sum_{i}&\left[t_{ss}^{(XM)}m_{(2)1is}^{\dagger}x_{(1)2is}\right.\\
&\left.+t_{X_s-M_p}^{(XM)}m_{(2)1ip_z}^{\dagger}x_{(1)2is}\right.\\
&\left.-t_{X_p-M_s}^{(XM)}m_{(2)1is}^{\dagger}x_{(1)2ip_z}\right.\\
&\left.+t^{(XM)}_{\pi}\sum_{\alpha=x,y}m_{(2)1ip_\alpha}^{\dagger}x_{(1)2ip_\alpha}\right.\\
&\left.-t^{(XM)}_{\sigma}m_{(2)1ip_z}^{\dagger}x_{(1)2ip_z}\right]+h.c.
\end{split}
\end{align}

The creation and annihilation operators now have additional indices for layers and sublayers, e.g. $x^{(\dagger)}_{(n)2is}$ annihilates (creates) an electron on layer $n$ ($n=1,2$), atom X, in sublayer 2, in orbital $s$. The sum over $i$ runs over all unit cells in the crystal. We denote inter-layer hopping parameters with a lowercase $t$. For the other M-X inter-layer interaction we have
\begin{widetext}
\begin{align}
\begin{split}
H_{M_1,X_{2}}=\sum_{<i,j>}&\left[t_{ss}^{(XX)}x_{(2)1js}^{\dagger}m_{(1)2is}+\sum_{\alpha}R^{M_{(1)2i}X_{(2)1j}}_{\alpha}\left(t_{Mp-Xs}^{(MX)}x_{(2)1js}^{\dagger}m_{(1)2ip_{\alpha}}-t_{Ms-Xp}^{(MX)}x_{(2)1jp_{\alpha}}^{\dagger}m_{(1)2is}\right)\right.\\
&\left.+\sum_{\alpha,\beta}\left(\delta_{\alpha\beta}t^{(MX)}_{\pi}-(t^{(MX)}_{\sigma}+t^{(MX)}_{\pi})R^{M_{(1)2i}X_{(2)1j}}_{\alpha}R^{M_{(1)2i}X_{(2)1j}}_{\beta}\right)x_{(2)1jp_{\beta}}^{\dagger}m_{(1)2ip_{\alpha}}\right]+h.c.
\end{split}
\end{align}
while X-X hoppings are included in the form
\begin{align}
\begin{split}
H_{X_1,X_{2}}=\sum_{<i,j>}&\left[t_{ss}^{(XX)}x_{(2)1js}^{\dagger}x_{(1)2is}+\sum_{\alpha}R^{X_{(1)2i}X_{(2)1j}}_{\alpha}t_{sp}^{(XX)}\left(x_{(2)1js}^{\dagger}x_{(1)2ip_{\alpha}}-x_{(2)1jp_{\alpha}}^{\dagger}x_{(1)2is}\right)\right.\\
&\left.+\sum_{\alpha,\beta}\left(\delta_{\alpha\beta}t^{(XX)}_{\pi}-(t^{(XX)}_{\sigma}+t^{(XX)}_{\pi})R^{X_{(1)2i}X_{(2)1j}}_{\alpha}R^{X_{(1)2i}X_{(2)1j}}_{\beta}\right)x_{(2)1jp_{\beta}}^{\dagger}x_{(1)2ip_{\alpha}}\right]+h.c.
\end{split}
\end{align}
\end{widetext} 

\noindent where the sums over $<i,j>$ are over nearest-neighboring X-X pairs and next-nearest-neighboring M-X pairs in adjacent sublayers. The inter-layer interactions included add 14 parameters to the model. When expressed in matrix form in a $\mathbf{k}$-space basis, the bilayer model gives a $32 \times 32$ matrix, which we diagonalize to obtain a set of $32$ bands.

To obtain the parameters, we fit the TB band structure to the DFT band structure of bilayer InSe obtained within the local density approximation. In the DFT calculation the monolayer geometry was kept fixed and the inter-layer distance set to 8.32 \AA, which corresponds to the experimentally known separation in $\gamma$-InSe \cite{Mudd2013}. We search for the ideal set of inter-layer hopping parameters to achieve the best least squares fit between the two band structures while keeping the inter-layer hopping parameters obtained in the monolayer model unchanged. In the monolayer we fitted the model to DFT data for 7 bands near the Fermi level. In the bilayer these bands split into subbands forming 14 bands in total, all of which are taken into account in the fitting procedure. As in the monolayer, we fit to the scissor-corrected DFT data, since the dependence of the optical transition matrix elements on $N$ is significantly affected by the size of the band gap - this is explored in detail in appendix \ref{SC_comp}.

\begin{figure}
 \centering
   \includegraphics[width=0.5\textwidth]{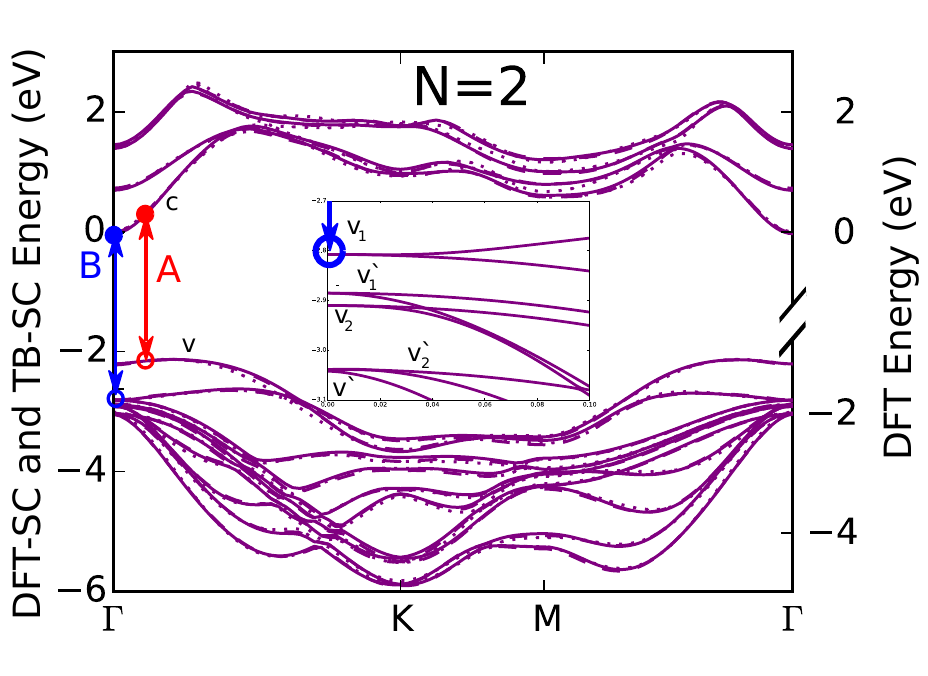}
\caption{(Color online) Band structures from DFT (dotted lines) and TB (solid lines with scissor correction and dashed lines without it) for bilayer $\gamma$-InSe. Zero of energy is set to the bottom of the conduction band. The left hand axis shows the scissor corrected DFT energies, the right hand the original energies. The inset shows a magnified view of the $\Gamma$-point region in the valence band.}
\label{fig_bilayer}
\end{figure}

The results of the fitting are presented alongside the DFT data for bilayer $\gamma$-InSe in Fig. \ref{fig_bilayer}, with the inter-layer TB parameters given in Table \ref{tab_multlayer}. We highlight the 8 bands derived from the monolayer bands $c$, $v$, $v_1$, and $v_2$; we label these $c^{\prime}$, $c$, $v$, $v_1$, $v_{1}^{\prime}$, $v_{2}$, $v^{\prime}$, and $v_{2}^{\prime}$. The zero of energy is set at the bottom of the conduction band. We provide the orbital decomposition of the $\Gamma$-point wave functions in Table \ref{Tab_2L_OrbDecompLDA}.

\begin{table*}
\caption{Relative weights on the valence $s$ and $p$ orbitals of In and Se atoms in
2-layer InSe at the $\Gamma$-point, for the bands labeled in Fig. \ref{fig_bilayer}. The equivalent contribution found in the scissor corrected TB model is given in square brackets.
Atoms are listed from bottom to top of 2L crystal. Band energies are provided relative to the
lowest conduction band edge (c). $E^{DFT}$ is the $\Gamma$-point energy value obtained using DFT
and  $E^{DFT-SC}=E^{DFT}-\delta E_{g}^{0K}$ is the value obtained after subtracting the scissor
correction.
The bands $v_1$, $v_{1}^{\prime}$, $v_{2}$, and $v_{2}^{\prime}$ are double degenerate at the $\Gamma$-point. The energies corresponding to $\hbar \omega _A$ and  $\hbar \omega _B$ are marked in bold.
\label{Tab_2L_OrbDecompLDA}}

\begin{center}
\begin{tabular}{lllllllll}

\hline
\hline

&$c^{\prime}$&$c$&$v$&$v_1$&$v_{1}^{\prime}$&$v_{2}$&$v^{\prime}$&$v_{2}^{\prime}$\\
\hline
$E^{DFT}$ (eV) &   0.69 &   0.00 &  -1.21 &  -1.81 &  -1.88 &  -1.91 &  -2.00 &  -2.03 \\
\rowcolor{lightgray} $E^{DFT-SC}$ (eV)&&&\textbf{-2.20}&\textbf{-2.80}&-2.87&-2.90&-2.99&-3.02\\
Se1 &  0.06[0.00]$s  $ &  0.05[0.00]$s  $ &  0.01[0.01]$s  $ &  0.06[0.07]$p_{x(y)}$ &  0.20[0.06]$p_{x(y)}$ &  0.15[0.36]$p_{x(y)}$ &  0.00[0.01]$s$           &  0.01[0.00]$p_{x(y)}$ \\
    &  0.03[0.07]$p_z$ &  0.11[0.15]$p_z$ &  0.18[0.23]$p_z$ &  &  &  &  0.16[0.14]$p_z$ &  \\
In1 &  0.10[0.05]$s  $ &  0.12[0.11]$s  $ &  0.03[0.07]$s  $ &  0.01[0.00]$p_{x(y)}$ &  0.04[0.01]$p_{x(y)}$ &  0.03[0.00]$p_{x(y)}$ &  0.00[0.03]$s  $ &  0.00[0.00]$p_{x(y)}$ \\
    &  0.06[0.05]$p_z$ &  0.00[0.07]$p_z$ &  0.06[0.01]$p_z$ &  &  &  &  0.07[0.02]$p_z$ &  \\
In2 &  0.06[0.09]$s  $ &  0.12[0.07]$s  $ &    0.00[0.03]$s$ &  0.03[0.01]$p_{x(y)}$ &  0.00[0.00]$p_{x(y)}$ &  0.01[0.00]$p_{x(y)}$ &  0.05[0.08]$s  $ &  0.03[0.01]$p_{x(y)}$ \\
    &  0.00[0.07]$p_z$ &  0.02[0.05]$p_z$ &  0.07[0.02]$p_z$ &  &  &  &  0.06[0.01]$p_z$ &  \\
Se2 &  0.01[0.00]$s  $ &  0.07[0.00]$s  $ &  0.01[0.01]$s  $ &  0.17[0.22]$p_{x(y)}$ &  0.01[0.01]$p_{x(y)}$ &  0.05[0.04]$p_{x(y)}$ &  0.02[0.01]$s  $ &  0.19[0.21]$p_{x(y)}$ \\
    &  0.16[0.17]$p_z$ &  0.03[0.07]$p_z$ &  0.14[0.15]$p_z$ &  &  &  &  0.15[0.19]$p_z$ &  \\
Se3 &  0.01[0.00]$s  $ &  0.06[0.00]$s  $ &  0.01[0.01]$s  $ &  0.16[0.19]$p_{x(y)}$ &  0.02[0.01]$p_{x(y)}$ &  0.04[0.03]$p_{x(y)}$ &  0.02[0.01]$s  $ &  0.20[0.25]$p_{x(y)}$ \\
    &  0.16[0.18]$p_z$ &  0.04[0.07]$p_z$ &  0.14[0.15]$p_z$ &  &  &  &  0.15[0.19]$p_z$ &  \\
In3 &  0.06[0.07]$s  $ &  0.11[0.08]$s  $ &   0.00[0.02]$s$   &  0.03[0.00]$p_{x(y)}$ &  0.00[0.00]$p_{x(y)}$ &  0.01[0.00]$p_{x(y)}$ &  0.05[0.10]$s  $ &  0.04[0.01]$p_{x(y)}$ \\
    &  0.01[0.03]$p_z$ &  0.02[0.07]$p_z$ &  0.08[0.02]$p_z$ &  &  &  &  0.06[0.01]$p_z$ &  \\
In4 &  0.11[0.07]$s  $ &  0.11[0.09]$s  $ &  0.04[0.07]$s  $ &  0.01[0.00]$p_{x(y)}$ &  0.03[0.00]$p_{x(y)}$ &  0.04[0.00]$p_{x(y)}$ &  0.00[0.04]$s  $ &  0.00[0.00]$p_{x(y)}$ \\
    &  0.07[0.08]$p_z$ &   0.00[0.05]$p_z$   &  0.06[0.00]$p_z$ &  &  &  &  0.07[0.02]$p_z$ &  \\
Se4 &  0.07[0.00]$s  $ &  0.05[0.00]$s  $ &  0.01[0.00]$s  $ &  0.04[0.01]$p_{x(y)}$ &  0.19[0.40]$p_{x(y)}$ &  0.17[0.06]$p_{x(y)}$ &  0.00[0.01]$s$           &  0.02[0.02]$p_{x(y)}$ \\
    &  0.04[0.07]$p_z$ &  0.10[0.14]$p_z$ &  0.19[0.22]$p_z$ &  &  &  &  0.16[0.15]$p_z$ &  \\

\hline
\hline

\end{tabular}
\end{center}
\end{table*}

\begin{table}
\centering
\caption{Inter-layer hopping parameters (eV) for scissor-corrected $\gamma$-InSe as defined in the Hamiltonian (Eq. \eqref{bilayer_hamiltonian}).}
\label{tab_multlayer}
\begin{tabular}{ccccc}
\hline\hline
$t_{ss}^{(XX)}$&$t_{sp}^{(XX)}$& $t_{\pi}^{(XX)}$&$t_{\sigma}^{(XX)}$\\
$-0.647$&$-0.626$&$-0.137$&$-0.830$\\ \hline
$t_{ss}^{(MX)}$&$t_{X_s-M_{pz}}^{(MX)}$&$t_{X_{pz}-M_s}^{(MX)}$&$t_{\pi}^{(MX)}$&$t_{\sigma}^{(MX)}$\\
$-0.397$&$0.112$&$-0.734$&$0.193$&$0.011$\\ \hline 
$t_{ss}^{(XM)}$&$t_{X_s-M_{pz}}^{(XM)}$&$t_{X_{pz}-M_s}^{(XM)}$&$t_{\pi}^{(XM)}$&$t_{\sigma}^{(XM)}$\\
-0.238&$0.042$&$-0.233$&$-0.398$&$0.450$\\ \hline \hline
\end{tabular}
\end{table}

\section{TB model for few-layer InSe: 2D bands and gaps}
\label{Few_layer_TBSC}
Applying our TB model to few-layer InSe requires the generalization of the bilayer model as follows. As in the bilayer, we consider the interactions between the nearest-neighboring X-X pairs and nearest and next-nearest M-X pairs on adjacent monolayers. This gives us a Hamiltonian of the form
\begin{equation}
H=\sum_{n=1}^N H_{n}+\sum_{n=1}^{N-1} H_{n,n+1}
\end{equation}

\noindent where $N$ is the total number of layers, $H_n$ is the monolayer Hamiltonian on layer $n$ as set out above, and $H_{n,n+1}$ takes into account inter-layer interactions between adjacent layers $n$ and $n+1$. It has the form
\begin{equation}
H_{n,n+1}=H_{X_n,M_{n+1}}+H_{M_n,X_{n+1}}+H_{X_n,X_{n+1}}.
\end{equation}

The vertical M-X contribution is

\begin{align}
\begin{split}
H_{X_n,M_{n+1}}=\sum_{i}&\left[t_{ss}^{(XM)}m_{(n+1)1is}^{\dagger}x_{(n)2is}\right.\\
&\left.+t_{X_s-M_p}^{(XM)}m_{(n+1)1ip_z}^{\dagger}x_{(n)2is}\right.\\
&\left.-t_{X_p-M_s}^{(XM)}m_{(n+1)1is}^{\dagger}x_{(n)2ip_z}\right.\\
&\left.+t^{(XM)}_{\pi}\sum_{\alpha=x,y}m_{(n+1)1ip_\alpha}^{\dagger}x_{(n)2ip_\alpha}\right.\\
&\left.-t^{(XM)}_{\sigma}m_{(n+1)1ip_z}^{\dagger}x_{(n)2ip_z}\right]+h.c.
\end{split}
\end{align}

For the other M-X inter-layer interaction we have
\begin{widetext}
\begin{align}
\begin{split}
H_{M_n,X_{n+1}}=\sum_{<i,j>}&\left[t_{ss}^{(XX)}x_{(n+1)1js}^{\dagger}m_{(n)2is}+\sum_{\alpha}R^{M_{(n)2i}X_{(n+1)1j}}_{\alpha}\left(t_{Mp-Xs}^{(MX)}x_{(n+1)1js}^{\dagger}m_{(n)2ip_{\alpha}}-t_{Ms-Xp}^{(MX)}x_{(n+1)1jp_{\alpha}}^{\dagger}m_{(n)2is}\right)\right.\\
&\left.+\sum_{\alpha,\beta}\left(\delta_{\alpha\beta}t^{(MX)}_{\pi}-(t^{(MX)}_{\sigma}+t^{(MX)}_{\pi})R^{M_{(n)2i}X_{(n+1)1j}}_{\alpha}R^{M_{(n)2i}X_{(n+1)1j}}_{\beta}\right)x_{(n+1)1jp_{\beta}}^{\dagger}m_{(n)2ip_{\alpha}}\right]+h.c.
\end{split}
\end{align}
while X-X hoppings are included in the form
\begin{align}
\begin{split}
H_{X_n,X_{n+1}}=\sum_{<i,j>}&\left[t_{ss}^{(XX)}x_{(n+1)1js}^{\dagger}x_{(n)2is}+\sum_{\alpha}R^{X_{(n)2i}X_{(n+1)1j}}_{\alpha}t_{sp}^{(XX)}\left(x_{(n+1)1js}^{\dagger}x_{(n)2ip_{\alpha}}-x_{(n+1)1jp_{\alpha}}^{\dagger}x_{(n)2is}\right)\right.\\
&\left.+\sum_{\alpha,\beta}\left(\delta_{\alpha\beta}t^{(XX)}_{\pi}-(t^{(XX)}_{\sigma}+t^{(XX)}_{\pi})R^{X_{(n)2i}X_{(n+1)1j}}_{\alpha}R^{X_{(n)2i}X_{(n+1)1j}}_{\beta}\right)x_{(n+1)1jp_{\beta}}^{\dagger}x_{(n)2ip_{\alpha}}\right]+h.c.
\end{split}
\end{align}
\end{widetext} 
where the sums over $<i,j>$ are over nearest-neighboring X-X pairs and next-nearest-neighboring M-X pairs in adjacent sublayers. When expressed in matrix form in a $\mathbf{k}$-space basis, the $N$-layer model gives a $16N \times 16N$ matrix, which we diagonalize to obtain a set of $16N$ bands. The matrix elements are given in the Appendix.

\begin{figure*}
 \centering
   \includegraphics[width=0.9\textwidth]{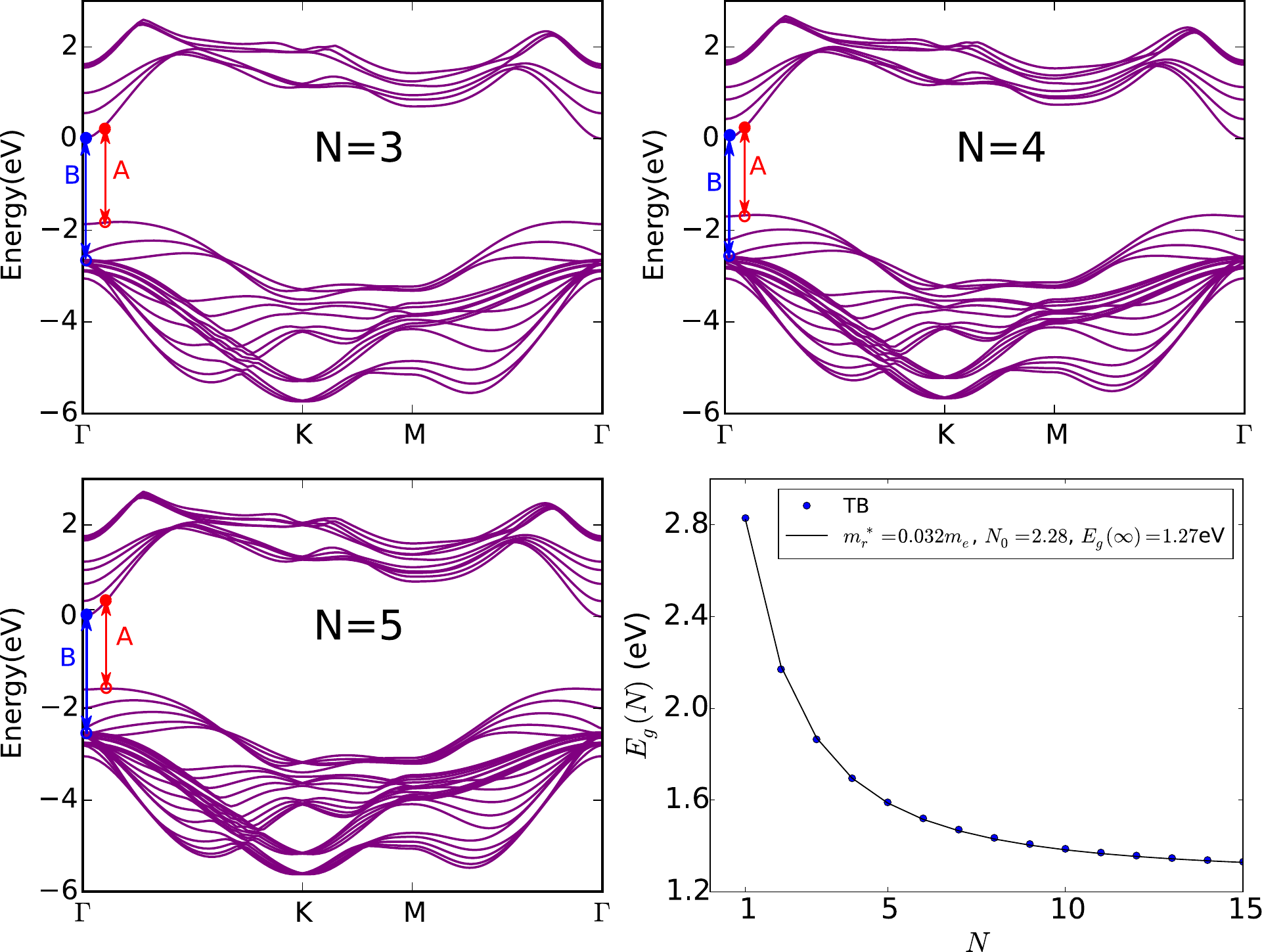}
\caption{(Color online) TB band structures for $N=$ 3, 4 and 5 layer $\gamma$-InSe. Zero of energy is set to the bottom of the conduction band. The bottom right panel shows the dependence of the vertical gap at the $\Gamma$ point relative to the bulk material on the number of layers in $N$-layer InSe, compared with a fit to the modified $k_z$ size-quantization gap model model, Eq. \eqref{eqn_gaps}.}
\label{fig_multlayer}
\end{figure*}

For the parameterization of the model we retain the hopping parameters from the bilayer model, corresponding to the approximation that the TB parameters will be the same for all values of $N$. Fig. \ref{fig_multlayer} shows the results of this extrapolation of the TB model to $N=$ 3, 4, and 5.

The bottom right panel of Fig. \ref{fig_multlayer} shows the vertical band gaps at $\Gamma$ according to the TB model at varying number of layers. If the band structure of bulk $\gamma-$InSe is available along $k_z$, one can extract the effective masses along the $k_z$ axis in the valence and conduction band, $m^{*}_{vz}$ and $m^{*}_{cz}$ respectively, and use these to apply the $k_z$ size-quantization gap model to approximate the expected gap for an $N$-layer structure, $E_g(N)$. This approximation strictly speaking only works for $N\gg 1$, but can be easily extended to few-layer materials using the following asymptotic formula

\begin{equation}
\label{eqn_gaps}
E_{g}(N)=E_g(\infty)+\frac{\hbar^2}{2m^{*}_r} \left \lbrack \frac{\pi}{a_z}\frac{1}{(N+N_0)}\right \rbrack ^2,
\end{equation}

\noindent with
$m^{*}_r=\dfrac{m^{*}_{cz} m^{*}_{vz}}{m^{*}_{cz}+m^{*}_{vz}}= 0.032m_e; N_0=2.28,$ obtained from fitting to vertical gaps from the TB model. The parameter $N_0$ is present to allow the model to retain its validity at a small number of layers, in which case the traditional effective mass model would need to be used with a general boundary condition, $\partial_z\psi=\alpha\psi$, to take into account that the wave function is pushed to the surface of the few-layer slabs, as we see in the wave functions calculated using the TB model. The behavior described by Eq. \eqref{eqn_gaps} is shown by a solid line in the right-hand lower panel in Fig. \ref{fig_multlayer}.

\section{4-band $\mathbf{k}\cdot\mathbf{p}$ theory for $N$- layer InSe and interband optical transitions}
\label{kp_section}
In the following we present a simple $\mathbf{k} \cdot \mathbf{p}$ model for the $c$, $v$, $v_1$, and $v_2$ bands in Fig. \ref{Energy_modelDFT_a}. In monolayer InSe these bands can be assigned the irreducible representations of point group $D_{3h}$ as seen in Fig. \ref{Energy_modelDFT_a}. The 4-band $\mathbf{k} \cdot \mathbf{p}$ Hamiltonian can be written as
\begin{widetext}
\begin{equation}
H = \left(
\begin{array}{cccc}
H_c   & \frac{\hbar \alpha _N e}{c m_e} \mathbf{k} \cdot \mathbf{A}+E_z d_z& \frac{\hbar \beta _N e}{c m_e} A& 0 \\
\frac{\hbar \alpha _N e}{cm_e} \mathbf{k} \cdot \mathbf{A}+E_z d_z& H_v& 0 & 0 \\
\frac{\hbar \beta _N e}{c m_e} A& 0 & H_{v_1}& 0 \\
0 & 0 & 0 & H_{v_2}\\
\end{array}
\right)
\label{eq_kdotp_4x4}
\end{equation}
\end{widetext}
\noindent where the diagonal components are the single band $\mathbf{k} \cdot \mathbf{p}$ Hamiltonians for the bands $c$, $v$, $v_1$, and $v_2$ as discussed below, while the off-diagonal components correspond to the interaction between the electrons and photons required to describe optical transitions between the bands $c$ and $v$, as well as between the bands $c$ and $v_1$. The one-band $\mathbf{k} \cdot \mathbf{p}$ description of the valence and conduction band is a straightforward polynomial expansion described below, while for bands $v_1$ and $v_2$ a suitable two-component Hamiltonian needs to be constructed that describes both branches in each band.

\begin{figure*}
 \centering
\includegraphics[width=0.90\textwidth]{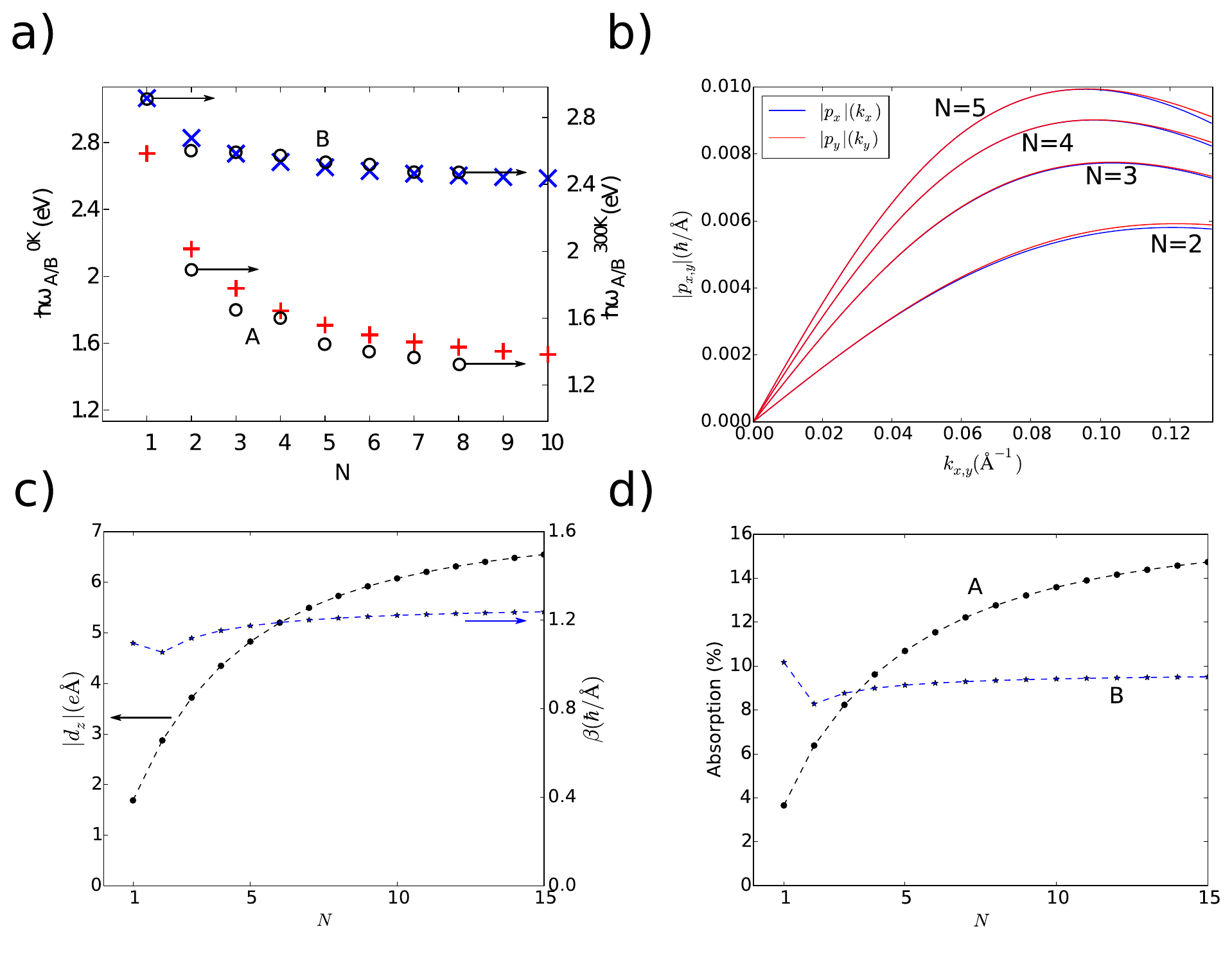}
\caption{(Color online) Transition properties of band edge excitations: (a) dependence of the energies of the principal transitions $A$ and $B$ on the number of layers $N$ in few-layer InSe according to DFT, with scissor corrections applied at $T=0K$ (left hand axis) and $T=300K$ (right hand axis), in comparison to experimental data measured at $T=300K$ taken from Ref. \onlinecite{Bandurin2016} (b) TB matrix elements for the $x$ and $y$ components of the interband momentum for the $A$--line for  $N=$ 2, 3, 4 and 5 layer $\gamma$-InSe (c) The $d_z$ matrix element and $\beta$, and (d) the absorption coefficient, $g_A(\theta)$ for incoming light arriving at the angle $\theta=\pi/4$ as a function of the number of layers $N$ and $g_B$ at $\theta=\pi/2$ for in-plane polarized light, obtained using the scissor-corrected TB model. Below we list the DFT-calculated energy gap $E_{A/B}^{DFT}$ and transition energies $\hbar \omega _{A/B}^{0K}$ and  $\hbar \omega _{A/B}^{300K}$ obtained using scissor correction at low and room temperature, the conduction band effective mass ($m_c$ in units of the free electron mass), the parameters $\alpha$ and $\beta$ for the $A$-- and $B$--transitions, and the values of $d_z$, for $N=1,2,3$.}
\begin{tabular}{cccccccc}
\hline
\hline
$N$&$E_{A/B}^{DFT}$(eV)& $\hbar \omega _{A/B}^{0K}$ (eV)& $\hbar \omega _{A/B}^{300K}$ (eV)&$m_c$ ($m_e$)&$\alpha $ &$\beta$ ($\hbar$/\AA)&$|d_z|$(e\AA)\\
\hline
1& 1.602 / 1.933& 2.734 / 3.066& 2.584 / 2.916& 0.188& 0.000& 1.096& 1.68\\
2& 1.031 / 1.695& 2.164 / 2.827& 2.014 / 2.677& 0.148& 0.082& 1.055& 2.87\\
3& 0.796 / 1.601& 1.929 / 2.734& 1.779 / 2.584& 0.132& 0.132& 1.119& 3.72\\
\hline
\hline
\end{tabular}
\label{fig_AB_energies_vs_N}
\end{figure*}

The bottom of the conduction band in 1L-InSe is quadratic in shape and can be described by the Hamiltonian

\begin{equation}
H_c=\hbar^2 k^2/2m_c + \gamma_c s_z k^3 \cos(3 \phi) +\kappa_c(N) (\mathbf{k} \times \mathbf{s})\mathbf{l_z}
\label{Ham_CB}
\end{equation}

\noindent where $\mathbf{k}$ is the electron wave-vector measured from the $\Gamma$-point,  $m_c$ is the effective mass at the conduction band minimum (listed in the caption of Fig. \ref{fig_AB_energies_vs_N}) and the second term in the Hamiltonian describes the spin-orbit splitting in the vicinity of the $\Gamma$-point, and $s_z=\pm 1/2$. The magnitude of the coupling constant is $\gamma_c=1.49(1)$ eV\AA$^3$ as found by fitting the energy splitting between the two spin components in the conduction band up to wave vectors less than 0.06 1/\AA. The spin-orbit splitting according to the local density approximation is presented in Fig. \ref{FigS4}. The lack of a splitting along the $\Gamma-M$ line is in agreement with the trigonal symmetry exhibited by the second term in the $\mathbf{k} \cdot \mathbf{p}$ Hamiltonian in Eq. \eqref{Ham_CB}. The last term in Eq. \eqref{Ham_CB} appears in $N$L-InSe only for $N>1$ and is present due to the breaking of the mirror-plane symmetry. The coefficient $\kappa_c$ is expected to depend on the number of layers.

The $N$ highest valence bands in $N$L-InSe are ``sombrero-shaped'' \cite{Zolyomi2014}. The highest valence band can be fitted around the $\Gamma$-point with an 8th order polynomial function as follows:

\begin{align}
\begin{split}
H_{v} (k,\phi)=&E_{v} + E_{2}k^{2} + E_{4}k^{4} +\\
&E_{6}k^{6} +E_{8}k^{8} + E_{6}^{\prime}k^6 \cos(6\phi) +\\
&\gamma_v s_z k^3 \cos(3 \phi) +\kappa_v(N) (\mathbf{k} \times \mathbf{s})\mathbf{l_z}
\label{8thOrder}
\end{split}
\end{align}

\noindent where the $E_{6}^{\prime}$ coefficient describes the hexagonal anisotropy. The fitted parameters are summarized in Table \ref{TabParams}. Note that the valence band takes the shape of an inverted sombrero which has been demonstrated to lead to a Lifshitz transition upon hole doping in the monolayer \cite{Zolyomi2014}. The sombrero shape and the associated Lifshitz transition persists with increasing $N$ but slowly vanishes as we approach the bulk limit. Accordingly, the critical carrier density required to achieve the transition decreases with increasing $N$, as shown in Table \ref{TabParams}.

The last two terms in Eq. \eqref{8thOrder} describe the spin-orbit splitting similar to Eq. \eqref{Ham_CB}. The magnitude of $\gamma_v$ according to a fit to the spin-orbit splitting in 1L-InSe for wave vectors less than 0.06 1/\AA~ is $\gamma_v=3.11(5)$ eV\AA$^3$. Note that the polynomial fit is valid in a 4-5 times larger range than the fit for the spin-orbit splitting.

\begin{table}
\caption{The parameters in Eq. \eqref{8thOrder} after fitting to the topmost valence band in the band structures of 1L, 2L, and 3L--InSe, with the zero of energy set to the valence band edge. The critical carrier density required to achieve the Lifshitz transition in the valence band is given as $n_{LT}$.
\label{TabParams}}
\begin{center}
\begin{tabular}{llll}
\hline
\hline
& 1L & 2L & 3L \\
\hline
$E_0$ (eV)& -0.078  &  -0.069 & -0.056\\
$E_2$ (eV\AA$^{2}$)& 2.915  &  4.767     & 5.318 \\
$E_4$ (eV\AA$^{4}$)& -38.057 & -106.817  &-163.540  \\
$E_6$ (eV\AA$^{6}$)& 206.551 &  896.029  & 1894.272\\
$E_6^{\prime}$ (eV\AA$^{6}$)& 3.050     & 5.658     & 6.500\\
$E_8$ (eV\AA$^{8}$)& -450.034 &-2982.703 &-8844.573 \\
$n_{LT}$ ($10^{13}$cm$^{-2}$) & 7.3 & 3.6 & 1.8 \\
\hline
\hline
\end{tabular}
\end{center}
\end{table}

The bands $v_1$ and $v_2$ are double degenerate at the $\Gamma$-point and can each be described by a two-component $\mathbf{k} \cdot \mathbf{p}$ model as described by the Hamiltonian

\begin{equation}
H_{v_{1(2)}}=E_{v_{1(2)}}+\frac{k^2}{2m}+\frac{k_x^2-k_y^2}{2m^{\prime}}\sigma_x +\frac{2k_x k_y}{2m^{\prime}}\sigma_y
\label{kdotp_v12}
\end{equation}

\noindent where $\sigma_{x(y)}$ are the Pauli matrices. This Hamiltonian transforms according to the $E$ irrep. of the symmetry group $C_{3v}$ (see Fig. \ref{Energy_modelDFT_a}).  Eq. \eqref{kdotp_v12} can be fitted to the DFT band structure to obtain the effective masses. In the band $v_1$ we obtain $m=0.31$ and $m^{\prime}=0.45$, and in the band $v_2$ we obtain $m=0.30$ and $m^{\prime}=0.45$ in units of $m_e$.

In Fig. \ref{fig_AB_energies_vs_N}a we show the energies of the $A$ and $B$ optical transitions at the $\Gamma$-point (energies $|E_v|$ and $E_{v1}$, respectively), where we apply the low-temperature scissor correction to the transition energies as discussed in Section \ref{model_vs_dft}. On the right hand side we show the same data with $T=300K$ scissor corrections for reference to room temperature measurements. The scissor corrected transition energies are summarized in the caption of Fig. \ref{fig_AB_energies_vs_N}.

To describe the coupling of the principal interband transition, A, between the conduction ($c$) and valence ($v)$ bands to an in-plane vector potential $\mathbf{A}$ carried by an incoming photon, $\mathbf{A\cdot P}$, we rely on the following formula for the interband momentum operator $\mathbf{P}$\cite{LewYanVoon1993}:

\begin{align}
\begin{split}
\label{eqn_p}
\mathbf{P}_{cv}(\mathbf{k})&=\bra{c}\mathbf{P}\ket{v}=\frac{m_e}{\hbar}\bra{c}\nabla_{\mathbf{k}}H\ket{v}\\
&=\frac{m_e}{\hbar}\sum_{o,o'}C^*_{c\mathbf{k}}(o)C_{v\mathbf{k}}(o')\nabla_{\mathbf{k}}H_{o,o'}(\mathbf{k}),
\end{split}
\end{align}

\noindent where the sum over $o,o'$ runs over the orbitals in the model, $C_{c(v)\mathbf{k}}(o)$ is the coefficient of the eigenfunction of the conduction(valence) band, orbital $o$, at $\mathbf{k}$, $H_{o,o'}=\bra{o}H\ket{o'}$, and $m_e$ is the free electron mass. We can therefore calculate the interband momentum matrix element using the above TB parameterization, with the matrix elements of the Hamiltonian and the eigenfunctions of the valence and conduction bands obtained directly from the TB model.

The TB matrix elements, seen in Fig. \ref{fig_AB_energies_vs_N}b, are linear at small $\mathbf{k}$ and the slope of this linear regime increases with an increasing number of layers. The latter observation is in line with the result that the momentum matrix element in the monolayer is zero. The finding that the matrix element is linear near the $\Gamma$-point allows the introduction of the dimensionless parameter $\alpha$ through the relation $\mathbf{P}_{cv}\simeq\hbar \alpha \mathbf{k}$ which is taken into account in Eq. \eqref{eq_kdotp_4x4}.

For coupling to the electric field associated with out-of-plane polarized light, we can also calculate the out-of-plane dipole matrix element $d_z=e\bra{c}\mathbf{z}\ket{v}$ between the valence and the conduction band. Since the crystal is finite in the $z$ direction we calculate the dipole matrix element directly as

\begin{equation}
d_z(\mathbf{k})=e\bra{c}\mathbf{z}\ket{v}(\mathbf{k})=e\sum_{o}C^*_{c\mathbf{k}}(o)C_{v\mathbf{k}}(o)\mathbf{z}(o)
\end{equation}

\noindent where the sum over $o$ is over all orbitals in the unit cell, 
$C_{c(v)\mathbf{k}}(o)$ is the coefficient of the conduction (valence) band 
eigenfunction for orbital $o$ at $\mathbf{k}$, and $\mathbf{z}(o)$ is the 
$z$-coordinate, w.r.t. the mean plane of the crystal, of the atom on which orbital 
$o$ sits.

The optical absorption coefficient for band edge absorption can be calculated from $d_z$ using Fermi's golden rule. A perturbation of $E_z d_z$ where $E_z$ is the electric field of the incoming photon, the rate of energy absorption in a material of dipole moment $d_z$ is

\begin{equation}
\Delta W = \hbar \omega \frac{2 \pi}{\hbar} g_S E_z^2 d_z^2 \sum_p \delta (\varepsilon_c(p)-\varepsilon_v(p)-\hbar \omega)
\end{equation}

\noindent where $\hbar \omega$ is the photon energy, $g_S=2$, and $\varepsilon_c(p)$ and $\varepsilon_v(p)$ are the band edge dispersions in the conduction and the valence band, respectively, as determined by $\mathbf{k}\cdot\mathbf{p}$ theory. The absorption coefficient $g(\theta)$ as a function of the angle $\theta$ between the incoming photon and the surface can be calculated simply by dividing $\Delta W$ by the absorbed energy, which is the flux of the Poynting vector over the visible area of the unit cell,

\begin{equation}
W_{tot} = A \sin (\theta) \frac{c}{4 \pi} E_z^2 \frac{1}{\cos(\theta)}
\end{equation}

\noindent where $A$ is the unit cell area. Evaluating this expression yields for the absorption at $\Gamma$

\begin{equation}
g_A(\theta) = 8 \pi \frac{e^2}{\hbar c} |d_z/e|^2 \frac{\hbar \omega m_c}{\hbar ^2} \cot (\theta) 
\end{equation}

\noindent where $m_c$ is the conduction band effective mass.

For coupling of the transition between bands $c$ and $v_1$, B, with in-plane polarized light, $\mathbf{A\cdot P}$, we evaluate $\mathbf{P}$ as
\begin{equation}
\mathbf{P}_{cv_1}=\frac{m_e}{\hbar}\sum_{o,o'}C^*_{c\mathbf{k}}(o)C_{v_1\mathbf{k}}(o')\nabla_{\mathbf{k}}H_{o,o'}(\mathbf{k}),
\end{equation}
from which we find how the transition B absorbs in-plane polarized light, at $\Gamma$,
\begin{equation}
g_B=8 \pi \frac{e^2}{\hbar c}\beta^2 \frac{m_c}{\hbar \omega m_e^2},
\end{equation}
\noindent where $\beta=|P_{cv(1)}|$ has finite values at $\Gamma$, listed in the caption to Fig. \ref{fig_AB_energies_vs_N} for $N=1,2,3$.
 In Fig. \ref{fig_AB_energies_vs_N}d we show the dependence of $g_A(\theta=\pi/4)$ and $g_B$ on the number of layers $N$; while $g_A$ exhibits a strong dependence on $N$, $g_B$ is almost constant.
\section{Conclusions}
\label{conclusions}
We have developed a TB model to describe monolayer and few-layer indium selenide which takes into account all $s$ and $p$ orbitals of constituent atoms. We have used first principles density functional theory to parametrize the model. We have found that:

\begin{itemize}

\item inclusion of $s$ and $p$ orbitals and hoppings to second-nearest-neighbors is sufficient to describe the energies of the bands near the band edge,

\item the interband optical matrix element obtained from our model exhibits a linear $\mathbf{k}$-dependence in agreement with DFT calculations, and

\item the matrix element vanishes in the monolayer due to symmetry.

\end{itemize}

We used the model to find the optical absorption coefficient in few-layer InSe: of the two principal optical transitions the absorption coefficient of the lower energy transition ($A$ line), corresponding to band edge absorption between the conduction and the valence band, slowly increases with the number of layers, while the absorption for the higher energy transition ($B$ line) saturates quickly to $\approx 10$ \%.

Also, we find that the conduction band electrons are relatively light ($m \propto 0.14-0.18 m_e$), in contrast to an almost flat dispersion of valence band holes near the $\Gamma$-point, which is found for up to $N \propto 6$. The latter property of the valence band suggests that this material may experience a phase transition due to many-body effects into either a ferromagnetic state as suggested for the similar material GaSe \cite{Cao2015}, or into a Peierls-type charge density wave due to a strong electron-phonon coupling \cite{ZCMF_InXGaX_EPC_unpub}.

The other members of the family of hexagonal III-VI semiconductors, such as GaSe\cite{Zolyomi2013}, have a similar crystal structure in the monolayer, and the TB model in section \ref{the_model} could be extended to cover these materials.  However, few-layer GaSe has a different consecutive layer arrangement, hence this will be covered in a future work\cite{stackings}.

\begin{acknowledgments}
The authors thank A. Patan\`e, L. Eaves, A. V. Tyurnina, D. A. Bandurin, A. K. Geim,  M. Potemski, and N. D. Drummond for discussions. This work made use of the facilities of N8 HPC provided and funded by the N8 consortium and EPSRC EP/K000225, the CSF cluster of the University of Manchester, and the High-End Computing cluster of Lancaster University. SJM acknowledges support from EPSRC CDT Graphene NOWNANO EP/L01548X. VF acknowledges support from ERC Synergy Grant Hetero2D, EPSRC EP/N010345, and Lloyd Register Foundation Nanotechnology grant. VZ and VF acknowledge support from the European Graphene Flagship Project.

\end{acknowledgments}
\appendix
\section{Monolayer Hamiltonian matrix elements}
\subsection{Mirror plane symmetry}
In a basis containing all $s$ and $p$ valence orbitals the Hamiltonian will be a $16 \times 16$ matrix. We can reduce the system to two $8 \times 8$ matrices by making use of the $\mathcal{M}_1$ symmetry of the crystal structure, which will require that the wavefunction be even or odd w.r.t. exchange of the two sublayers. We therefore construct a new basis from even and odd combinations of our orbitals:

\begin{align}
\begin{split}
{m_{is}^{(\pm)}}^{\dagger}&=\frac{1}{\sqrt{2}}(m_{1is}^{\dagger}\pm m_{2is}^{\dagger}),\\
{x_{is}^{(\pm)}}^{\dagger}&=\frac{1}{\sqrt{2}}(x_{1is}^{\dagger}\pm x_{2is}^{\dagger}),\\
{m_{ip\alpha}^{(\pm)}}^{\dagger}&=\frac{1}{\sqrt{2}}(m_{1ip\alpha}^{\dagger}\pm m_{2ip\alpha}^{\dagger}),\\
{x_{ip\alpha}^{(\pm)}}^{\dagger}&=\frac{1}{\sqrt{2}}(x_{1ip\alpha}^{\dagger}\pm x_{2ip\alpha}^{\dagger}),\\
{m_{ipz}^{(\pm)}}^{\dagger}&=\frac{1}{\sqrt{2}}(m_{1ipz}^{\dagger}\mp m_{2ipz}^{\dagger}),\\
{x_{ipz}^{(\pm)}}^{\dagger}&=\frac{1}{\sqrt{2}}(x_{1ipz}^{\dagger}\mp x_{2ipz}^{\dagger}),
\end{split}
\end{align}

\noindent where $\alpha=x,y$ and $p_z$ orbitals have an extra $(-)$ sign on the bottom sublayer contribution as the direction of $p_z$ is reversed under $\mathcal{M}_1$. A matrix constructed in the above basis will be block-diagonal, as mixing between even and odd states would break $\mathcal{M}_1$ symmetry.

\subsection{Representation}

To calculate our Hamiltonian we express it in a $\mathbf{k}$-space basis, constructing a matrix with elements $H_{ab}=a_{\mathbf{k}}Hb_{\mathbf{k}}^{\dagger}$ where $a_{\mathbf{k}}^{(\dagger)}$ and $b_{\mathbf{k}}^{(\dagger)}$ are the annihilation (creation) operators for the orbitals in our basis at wave vector $\mathbf{k}$ in the Brillouin zone. The operators $a_{\mathbf{k}}^{(\dagger)}$ can be expressed with the real space annihilation (creation) operators $a_{i}^{(\dagger)}$ as

\begin{equation}
a_{\mathbf{k}}=\frac{1}{\sqrt{N_{latt}}}\sum_{i}e^{i\mathbf{k}\cdot \mathbf{R_{i}}}a_{i}
\end{equation}
\noindent where $R_i$ is the position of the real space orbital $a_i$, and $N_{latt}$ is the number of lattice sites.

Relying on the symmetry adapted basis, we represent our Hamiltonian as two $8 \times 8$ matrices in the $\mathbf{k}$-space basis with elements of the form:
\begin{align}
\begin{split}
H^{(\pm)}_{ab}&=a_{\mathbf{k}}^{(\pm)}Hb_{\mathbf{k}}^{(\pm)\dagger}.
\end{split}
\end{align}

Now substituting in the original forms of the even and odd basis, we get for two orbitals where neither are $p_z$
\begin{align}
\begin{split}
H^{(\pm)}_{ab}&=\frac{1}{2}[a_{1\mathbf{k}}\pm a_{2\mathbf{k}}]H[b_{1\mathbf{k}}^{\dagger}\pm b_{2\mathbf{k}}^{\dagger}]\\
&=\frac{1}{2}\left[a_{1\mathbf{k}}Hb_{1\mathbf{k}}^{\dagger}+a_{2\mathbf{k}}Hb_{2\mathbf{k}}^{\dagger}\pm a_{1\mathbf{k}}Hb_{2\mathbf{k}}^{\dagger}\pm  a_{2\mathbf{k}}Hb_{1\mathbf{k}}^{\dagger}\right].
\end{split}
\end{align}

As the system and the Hamiltonian are even under $\mathcal{M}_1$ we can observe that
\begin{align}
\begin{split}
a_{1\mathbf{k}}Hb_{1\mathbf{k}}^{\dagger}&=a_{2\mathbf{k}}Hb_{2\mathbf{k}}^{\dagger}\\
a_{1\mathbf{k}}Hb_{2\mathbf{k}}^{\dagger}&=a_{2\mathbf{k}}Hb_{1\mathbf{k}}^{\dagger}
\end{split}
\end{align}
and hence
\begin{align}
\begin{split}
H^{(\pm)}_{ab}=a_{1\mathbf{k}}Hb_{1\mathbf{k}}^{\dagger}\pm a_{2\mathbf{k}}Hb_{1\mathbf{k}}^{\dagger}.
\end{split}
\end{align}

In the case where orbital $a$ is $p_z$ we have
\begin{align}
\begin{split}
H^{(\pm)}_{ab}=a_{1\mathbf{k}}Hb_{1\mathbf{k}}^{\dagger}\mp a_{2\mathbf{k}}Hb_{1\mathbf{k}}^{\dagger}.
\end{split}
\end{align}

We therefore need only consider $H_0$, $H_{11}$ and $H_{12}$ in the calculation of our Hamiltonian matrix. We can then diagonalize the even and odd parts of the Hamiltonian separately, obtaining a set of 8 bands for each. The matrices have the form
\begin{widetext}
\begin{equation}
H^{(\pm)}=
\left[
\begin{array}{cccc|cccc}
H^{(\pm)}_{M_s,M_s}&H^{(\pm)}_{M_s,M_{px}}&H^{(\pm)}_{M_s,M_{py}}&H^{(\pm)}_{M_s,M_{pz}} &H^{(\pm)}_{M_s,X_s} & H^{(\pm)}_{M_s,X_{px}}& H^{(\pm)}_{M_s,X_{py}}& H^{(\pm)}_{M_s,X_{pz}}\\
H^{(\pm)*}_{M_s,M_{px}}&H^{(\pm)}_{M_{px},M_{px}}&H^{(\pm)}_{M_{px},M_{py}}&H^{(\pm)}_{M_{px},M_{pz}} &H^{(\pm)}_{M_{px},X_s} & H^{(\pm)}_{M_{px},X_{px}}& H^{(\pm)}_{M_{px},X_{py}}& H^{(\pm)}_{M_{px},X_{pz}}\\
H^{(\pm)*}_{M_s,M_{py}}&H^{(\pm)*}_{M_{px},M_{py}}&H^{(\pm)}_{M_{py},M_{py}}&H^{(\pm)}_{M_{py},M_{pz}} &H^{(\pm)}_{M_{py},X_s} & H^{(\pm)}_{M_{py},X_{px}}& H^{(\pm)}_{M_{py},X_{py}}& H^{(\pm)}_{M_{py},X_{pz}}\\
H^{(\pm)*}_{M_s,M_{pz}}&H^{(\pm)*}_{M_{px},M_{pz}}&H^{(\pm)*}_{M_{py},M_{pz}}&H^{(\pm)}_{M_{pz},M_{pz}} &H^{(\pm)}_{M_{pz},X_s} & H^{(\pm)}_{M_{pz},X_{px}}& H^{(\pm)}_{M_{pz},X_{py}}& H^{(\pm)}_{M_{pz},X_{pz}}\\ \hline
H^{(\pm)*}_{M_s,X_s}&H^{(\pm)*}_{M_{px},X_s}&H^{(\pm)*}_{M_{py},X_s}&H^{(\pm)*}_{M_{pz},X_s} &H^{(\pm)}_{X_s,X_s} & H^{(\pm)}_{X_s,X_{px}}& H^{(\pm)}_{X_s,X_{py}}& H^{(\pm)}_{X_s,X_{pz}}\\
H^{(\pm)*}_{M_s,X_{px}}&H^{(\pm)*}_{M_{px},X_{px}}&H^{(\pm)*}_{M_{py},X_{px}}&H^{(\pm)*}_{M_{pz},X_{px}} &H^{(\pm)*}_{X_s,X_{px}} & H^{(\pm)}_{X_{px},X_{px}}& H^{(\pm)}_{X_{px},X_{py}}& H^{(\pm)}_{X_{px},X_{pz}}\\
H^{(\pm)*}_{M_s,X_{py}}&H^{(\pm)*}_{M_{px},X_{py}}&H^{(\pm)*}_{M_{py},X_{py}}&H^{(\pm)*}_{M_{pz},X_{py}} &H^{(\pm)*}_{X_s,X_{py}} & H^{(\pm)*}_{X_{px},X_{py}}& H^{(\pm)}_{X_{py},X_{py}}& H^{(\pm)}_{X_{py},X_{pz}}\\
H^{(\pm)*}_{M_s,X_{pz}}&H^{(\pm)*}_{M_{px},X_{pz}}&H^{(\pm)*}_{M_{py},X_{pz}}&H^{(\pm)*}_{M_{pz},X_{pz}} &H^{(\pm)*}_{X_s,X_{pz}} & H^{(\pm)*}_{X_{px},X_{pz}}& H^{(\pm)*}_{X_{py},X_{pz}}& H^{(\pm)}_{X_{pz},X_{pz}}\\
\end{array}
\right]
\end{equation}
\end{widetext}
The elements are calculated as set out above. For the calculation of the Bloch phase factors we can reduce the hopping vectors to three sets, as the Brillouin zone is two-dimensional. These are 
for M-X hoppings
\begin{equation}
\boldsymbol{r}_1=\left[
\begin{array}{c}
0\\
-\frac{a}{\sqrt{3}}\\
\end{array}
\right],
\boldsymbol{r}_2=\left[
\begin{array}{c}
-\frac{a}{2}\\
\frac{a}{2\sqrt{3}}\\
\end{array}
\right],
\boldsymbol{r}_3=\left[
\begin{array}{c}
\frac{a}{2}\\
\frac{a}{2\sqrt{3}}\\
\end{array}
\right],
\end{equation}
and for M-M and X-X hoppings between ions in the same sublayer ($\mathbf{k}\cdot\mathbf{x}=0$ for M-M hopping between the two sublayers, where the ions are directly above each other) there are six such vectors 
\begin{equation}
\boldsymbol{r}_{4,7}=\pm\mathbf{a}_1,
\boldsymbol{r}_{5,8}=\pm\mathbf{a}_2,
\boldsymbol{r}_{6,9}=\pm(\mathbf{a}_1+\mathbf{a}_2).
\end{equation}
For next-nearest M-X pairs ($T^{(3)}$) we have
\begin{equation}
\boldsymbol{r}_{10}=\left[
\begin{array}{c}
0\\
\frac{2a}{\sqrt{3}}\\
\end{array}
\right],
\boldsymbol{r}_{11}=\left[
\begin{array}{c}
-a\\
-\frac{a}{\sqrt{3}}\\
\end{array}
\right],
\boldsymbol{r}_{12}=\left[
\begin{array}{c}
a\\
-\frac{a}{\sqrt{3}}\\
\end{array}
\right].
\end{equation}
The  $\mathbf{k}$-dependence then appears in the model through combinations of the Bloch phase factors calculated using these vectors
\begin{align}
\begin{split}
f_1&=e^{i\mathbf{k}\cdot\boldsymbol{r}_1}+e^{i\mathbf{k}\cdot\boldsymbol{r}_2}+e^{i\mathbf{k}\cdot\boldsymbol{r}_3},\\
f_{2}&=e^{i\mathbf{k}\cdot\boldsymbol{r}_2}-e^{i\mathbf{k}\cdot\boldsymbol{r}_3},\\
f_{3}&=e^{i\mathbf{k}\cdot\boldsymbol{r}_2}+e^{i\mathbf{k}\cdot\boldsymbol{r}_3},\\
f_4&=2e^{i\mathbf{k}\cdot\boldsymbol{r}_1}-e^{i\mathbf{k}\cdot\boldsymbol{r}_2}-e^{i\mathbf{k}\cdot\boldsymbol{r}_3},\\ 
f_5&=4e^{i\mathbf{k}\cdot\boldsymbol{r}_1}+e^{i\mathbf{k}\cdot\boldsymbol{r}_2}+e^{i\mathbf{k}\cdot\boldsymbol{r}_3},\\
f_6&=2\left[\cos(\mathbf{k}\cdot\boldsymbol{r}_4)+\cos(\mathbf{k}\cdot\boldsymbol{r}_5)+\cos(\mathbf{k}\cdot\boldsymbol{r}_6)\right],\\
f_7&=2\left[\cos(\mathbf{k}\cdot\boldsymbol{r}_4)+\cos(\mathbf{k}\cdot\boldsymbol{r}_5)+4\cos(\mathbf{k}\cdot\boldsymbol{r}_6)\right],\\
f_8&=2\left[\cos(\mathbf{k}\cdot\boldsymbol{r}_4)+\cos(\mathbf{k}\cdot\boldsymbol{r}_5)\right],\\
f_9&=2\left[\cos(\mathbf{k}\cdot\boldsymbol{r}_4)-\cos(\mathbf{k}\cdot\boldsymbol{r}_5)\right],\\
f_{10}&=2i\left[\sin(\mathbf{k}\cdot\boldsymbol{r}_4)+\sin(\mathbf{k}\cdot\boldsymbol{r}_5)+2\sin(\mathbf{k}\cdot\boldsymbol{r}_6)\right],\\
f_{11}&=2i\left[\sin(\mathbf{k}\cdot\boldsymbol{r}_4)-\sin(\mathbf{k}\cdot\boldsymbol{r}_5)\right],\\
f_{12}&=e^{i\mathbf{k}\cdot\boldsymbol{r}_{10}}+e^{i\mathbf{k}\cdot\boldsymbol{r}_{11}}+e^{i\mathbf{k}\cdot\boldsymbol{r}_{12}},\\
f_{13}&=e^{i\mathbf{k}\cdot\boldsymbol{r}_{11}}-e^{i\mathbf{k}\cdot\boldsymbol{r}_{12}},\\
f_{14}&=e^{i\mathbf{k}\cdot\boldsymbol{r}_{11}}+e^{i\mathbf{k}\cdot\boldsymbol{r}_{12}},\\
f_{15}&=2e^{i\mathbf{k}\cdot\boldsymbol{r}_{10}}-e^{i\mathbf{k}\cdot\boldsymbol{r}_{11}}-e^{i\mathbf{k}\cdot\boldsymbol{r}_{12}},\\ 
f_{16}&=4e^{i\mathbf{k}\cdot\boldsymbol{r}_{10}}+e^{i\mathbf{k}\cdot\boldsymbol{r}_{11}}+e^{i\mathbf{k}\cdot\boldsymbol{r}_{12}}.
\end{split}
\end{align}

The symbols $L_1$ and $L_2$ are the magnitudes of the hopping vectors for M-X intra- and inter-sublayer hoppings, respectively, and are given by
\begin{align}
\begin{split}
L_1&=\sqrt{\frac{a^2}{3}+\frac{(d_{XX}-d_{MM})^2}{4}},\\
L_2&=\sqrt{\frac{a^2}{3}+\frac{(d_{XX}+d_{MM})^2}{4}},\\
L_3&=\sqrt{a^2+d_{MM}^2},\\
L_4&=\sqrt{\frac{4a^2}{3}+\frac{(d_{XX}-d_{MM})^2}{4}}.
\end{split}
\end{align}
The matrix elements are:
\begin{widetext}
\subsubsection{Diagonal elements}
\vspace{-20pt}
\begin{align*}
H^{(\pm)}_{M_s,M_s}=\varepsilon_{Ms}\pm T_{ss}'^{(1)}+f_6\left[T_{ss}^{(2M)}\pm T_{ss}'^{(3)}\right]
\end{align*}
\begin{align*}
H^{(\pm)}_{M_{px},M_{px}}&=\varepsilon_{M_{px}}\pm T_{\pi}'^{(1)}+f_6\left[T_{\pi}^{(2M)}\pm T_{\pi}'^{(3)}\right]-\frac{f_7}{4}\left[T_{\pi}^{(2M)}+T_{\sigma}^{(2M)}\pm\frac{a^2}{L_3^2}\left(T_{\pi}'^{(3)}+T_{\sigma}'^{(3)}\right)\right]
\end{align*}
\begin{align*}
H^{(\pm)}_{M_{py},M_{py}}&=\varepsilon_{M_{px}}\pm T_{\pi}'^{(1)}+f_6\left[T_{\pi}^{(2M)}\pm T_{\pi}'^{(3)}\right]-\frac{3f_8}{4}\left[T_{\pi}^{(2M)}+T_{\sigma}^{(2M)}\pm\frac{a^2}{L_3^2}\left(T_{\pi}'^{(3)}+T_{\sigma}'^{(3)}\right)\right]
\end{align*}
\begin{align*}
H^{(\pm)}_{M_{pz},M_{pz}}&=\varepsilon_{M_{pz}}\pm T_{\sigma}'^{(1)}+f_6\left[T_{\pi}^{(2M)}\pm \left(T_{\pi}'^{(3)}-\frac{d_{MM}^2}{L_3^2}\left[T_{\pi}'^{(3)}+T_{\sigma}'^{(3)}\right]\right)\right]
\end{align*}
\begin{align*}
H^{(\pm)}_{X_s,X_s}&=\varepsilon_{Xs}+f_6T^{(2X)}_{ss}
\\
H^{(\pm)}_{X_{px},X_{px}}&=\varepsilon_{X_{px}}+f_6T_{\pi}^{(2X)}-\frac{f_7}{4}(T_{\pi}^{(2X)}+T_{\sigma}^{(2X)})
\\
H^{(\pm)}_{X_{py},X_{py}}&=\varepsilon_{X_{px}}+f_6T_{\pi}^{(2X)}-\frac{3f_8}{4}(T_{\pi}^{(2X)}+T_{\sigma}^{(2X)})
\\
H^{(\pm)}_{X_{pz},X_{pz}}&=\varepsilon_{X_{pz}}+f_6T_{\pi}^{(2X)}
\end{align*}
\subsubsection{M-X Off-diagonal elements}
\begin{align*}
H^{(\pm)}_{M_s,X_s}&=f_1(T_{ss}^{(1)}\pm T_{ss}'^{(2)})+f_{12}T_{ss}^{(3)}
\end{align*}
\begin{align*}
H^{(\pm)}_{M_s,X_{px}}&=-\frac{f_2a}{2}\left[\frac{T_{M_{s}-X_{p}}^{(1)}}{L_1}\pm\frac{T_{M_{s}-X_{p}}'^{(2)}}{L_2}\right]+\frac{f_{13}a}{L_4}T_{M_{s}-X_{p}}^{(3)}
\end{align*}
\begin{align*}
H^{(\pm)}_{M_s,X_{py}}&=\frac{f_4a}{2\sqrt{3}}\left[\frac{T_{M_{s}-X_{p}}^{(1)}}{L_1}\pm\frac{T_{M_{s}-X_{p}}'^{(2)}}{L_2}\right]-\frac{f_{15}a}{L_4\sqrt{3}}T_{M_{s}-X_{p}}^{(3)}
\end{align*}
\begin{align*}
H^{(\pm)}_{M_s,X_{pz}}&=-\frac{f_1}{2}\left[\frac{(d_{xx}-d_{MM})T_{M_{s}-X_{p}}^{(1)}}{L_1}\pm\frac{(d_{XX}+d_{MM})T_{M_{s}-X_{p}}'^{(2)}}{L_2}\right]-\frac{f_{12}(d_{XX}-d_{MM})}{2L_4}T_{M_{s}-X_{p}}^{(3)}
\end{align*}
\begin{align*}
H^{(\pm)}_{M_{px},X_{s}}&=\frac{f_2a}{2}\left[\frac{T_{M_{p}-X_{s}}^{(1)}}{L_1}\pm\frac{T_{M_{p}-X_{s}}'^{(2)}}{L_2}\right]-\frac{f_{13}a}{L_4}T_{M_{p}-X_{s}}^{(3)}
\end{align*}
\begin{align*}
H^{(\pm)}_{M_{py},X_{s}}&=-\frac{f_4a}{2\sqrt{3}}\left[\frac{T_{M_{p}-X_{s}}^{(1)}}{L_1}\pm\frac{T_{M_{p}-X_{s}}'^{(2)}}{L_2}\right]+\frac{f_{15}a}{L_4\sqrt{3}}T_{M_{p}-X_{s}}^{(3)}
\end{align*}
\begin{align*}
H^{(\pm)}_{M_{pz},X_{s}}&=\frac{f_1}{2}\left[\frac{(d_{XX}-d_{MM})T_{M_{p}-X_{s}}^{(1)}}{L_1}\mp\frac{(d_{XX}+d_{MM})T_{M_{p}-X_{s}}'^{(2)}}{L_2}\right]+\frac{f_{12}(d_{XX}-d_{MM})}{2L_4}T_{M_{p}-X_{s}}^{(3)}
\end{align*}
\begin{align*}
H^{(\pm)}_{M_{px},X_{px}}&=f_1(T_{\pi}^{(1)}\pm T_{\pi}'^{(2)})-\frac{f_3a^2}{4}\left[\frac{T_{\pi}^{(1)}+T_{\sigma}^{(1)}}{L_1^2}\pm\frac{T_{\pi}'^{(2)}+T_{\sigma}'^{(2)}}{L_2^2}\right]+f_{12}T_{\pi}^{(3)}-\left[\frac{f_{14}a^2}{L_4^2}\left(T_{\pi}^{(3)}+T_{\sigma}^{(3)}\right)\right]
\end{align*}
\begin{align*}
H^{(\pm)}_{M_{py},X_{py}}&=f_1(T_{\pi}^{(1)}\pm T_{\pi}'^{(2)})-\frac{f_5a^2}{12}\left[\frac{T_{\pi}^{(1)}+T_{\sigma}^{(1)}}{L_1^2}\pm\frac{T_{\pi}'^{(2)}+T_{\sigma}'^{(2)}}{L_2^2}\right]+f_{12}T_{\pi}^{(3)}-\left[\frac{f_{16}a^2}{L_4^2}\left(T_{\pi}^{(3)}+T_{\sigma}^{(3)}\right)\right]
\end{align*}
\begin{align*}
H^{(\pm)}_{M_{pz},X_{pz}}&=f_1\left[T_{\pi}^{(1)}\mp T_{\pi}'^{(2)}-\left(\frac{d_{XX}-d_{MM}}{2L_1}\right)^2(T_{\pi}^{(1)}+T_{\sigma}^{(1)})\pm \left(\frac{d_{XX}+d_{MM}}{2L_2}\right)^2(T_{\pi}'^{(2)}+T_{\sigma}'^{(2)}) \right]\\
&+f_{12}\left[T_{\pi}^{(3)}-\left(\frac{d_{XX}-d_{MM}}{2L_4}\right)^2(T_{\pi}^{(3)}+T_{\sigma}^{(3)})\right]
\end{align*}
\begin{align*}
H^{(\pm)}_{M_{px},X_{py}}&=H_{M_{py},X_{px}}=-\frac{f_2a^2}{4\sqrt{3}}\left[\frac{T_{\pi}^{(1)}+T_{\sigma}^{(1)}}{L_1^2}\pm\frac{T_{\pi}'^{(2)}+T_{\sigma}'^{(2)}}{L_2^2}\right]-\frac{f_{13}a^2}{L_4^2\sqrt{3}}\left(T_{\pi}^{(3)}+T_{\sigma}^{(3)}\right)
\end{align*}
\begin{align*}
H^{(\pm)}_{M_{px},X_{pz}}&=-\frac{f_2a}{4}\left[\frac{(d_{XX}-d_{MM})(T_{\pi}^{(1)}+T_{\sigma}^{(1)})}{L_1^2}\pm\frac{(d_{XX}+d_{MM})(T_{\pi}'^{(2)}+T_{\sigma}'^{(2)})}{L_2^2}\right]+\frac{f_{13}a(d_{XX}-d_{MM})}{2L_4^2}\left(T_{\pi}^{(3)}+T_{\sigma}^{(3)}\right)
\end{align*}
\begin{align*}
H_{M_{pz},X_{px}}&=-\frac{f_2a}{4}\left[\frac{(d_{XX}-d_{MM})(T_{\pi}^{(1)}+T_{\sigma}^{(1)})}{L_1^2}\mp\frac{(d_{XX}+d_{MM})(T_{\pi}'^{(2)}+T_{\sigma}'^{(2)})}{L_2^2}\right]+\frac{f_{13}a(d_{XX}-d_{MM})}{2L_4^2}\left(T_{\pi}^{(3)}+T_{\sigma}^{(3)}\right)
\end{align*}
\begin{align*}
H^{(\pm)}_{M_{py},X_{pz}}&=\frac{f_4a}{4\sqrt{3}}\left[\frac{(d_{XX}-d_{MM})(T_{\pi}^{(1)}+T_{\sigma}^{(1)})}{L_1^2}\pm\frac{(d_{XX}+d_{MM})(T_{\pi}'^{(2)}+T_{\sigma}'^{(2)})}{L_2^2}\right]-\frac{f_{15}a(d_{XX}-d_{MM})}{2\sqrt{3}L_4^2}\left(T_{\pi}^{(3)}+T_{\sigma}^{(3)}\right)
\end{align*}
\begin{align*}
H_{M_{pz},X_{py}}&=\frac{f_4a}{4\sqrt{3}}\left[\frac{(d_{XX}-d_{MM})(T_{\pi}^{(1)}+T_{\sigma}^{(1)})}{L_1^2}\mp\frac{(d_{XX}+d_{MM})(T_{\pi}'^{(2)}+T_{\sigma}'^{(2)})}{L_2^2}\right]-\frac{f_{15}a(d_{XX}-d_{MM})}{2\sqrt{3}L_4^2}\left(T_{\pi}^{(3)}+T_{\sigma}^{(3)}\right)
\end{align*}
\end{widetext}
\subsubsection{M-M, X-X off-diagonal elements}
\begin{align*}
H^{(\pm)}_{M_s,M_{px}}=-\frac{f_{10}}{2}\left[T_{sp}^{(2M)}\pm\frac{a}{L_3}T_{sp}^{(3)}\right]
\end{align*}
\begin{align*}
H^{(\pm)}_{M_s,M_{py}}=-\frac{f_{11}\sqrt{3}}{2}\left[T_{sp}^{(2M)}\pm\frac{a}{L_3}T_{sp}^{(3)}\right]
\end{align*}
\begin{align*}
H^{(\pm)}_{M_s,M_{pz}}=\mp\left[T_{sp}'^{(1M)}+f_6\frac{d_{MM}}{L_3}T_{sp}^{(3)}\right]
\end{align*}
\begin{align*}
H^{(\pm)}_{M_{px},M_{py}}&=-\frac{f_9\sqrt{3}}{4}\left[T_{\pi}^{(2M)}+T_{\sigma}^{(2M)}\pm\left(T_{\pi}^{(3)}+T_{\sigma}^{(3)}\right)\right]
\\
H^{(\pm)}_{M_{px},M_{pz}}&=\mp \frac{f_{10}d_{MM}a}{2L_3^2}\left[T_{\pi}^{(3)}+T_{\sigma}^{(3)}\right]\\
H^{(\pm)}_{M_{py},M_{pz}}&=\mp \frac{f_{9}d_{MM}a}{2L_3^2}\left[T_{\pi}^{(3)}+T_{\sigma}^{(3)}\right]
\end{align*}
\begin{align*}
H^{(\pm)}_{X_s,X_{px}}&=-\frac{f_{10}}{2}T_{sp}^{(2X)}
\\
H^{(\pm)}_{X_s,X_{py}}&=-\frac{f_{11}\sqrt{3}}{2}T_{sp}^{(2X)}
\\
H^{(\pm)}_{X_{px},X_{py}}&=-\frac{f_{9}\sqrt{3}}{4}(T_{\pi}^{(2X)}+T_{\sigma}^{(2X)})
\\
H^{(\pm)}_{X_s,X_{pz}}&=H^{(\pm)}_{X_{px},X_{pz}}=H^{(\pm)}_{X_{py},X_{pz}}=0
\end{align*}
\section{Inter-layer Hamiltonian matrix elements}
The Hamiltonian for bilayer $\gamma$-InSe is expressed in the form
\begin{equation}
H=
\left[
\begin{array}{cc}
H_{1\mathrm{L}}&H_c\\
H_c^{\dagger}&H_{1\mathrm{L}}
\end{array}
\right]
\end{equation}
where $H_{1\mathrm{L}}$ is the $16 \times 16$ monolayer Hamiltonian, expressed in the original atomic basis ($M_1$, $M_2$, $X_1$, $X_2$, as opposed to the even/odd basis used above), and $H_c$ includes the inter-layer interactions. We write $H_c$ as
\begin{equation}
H_c=
\left[
\begin{array}{cccc}
0&0&0&0\\
0&0&H_{M_{(1)2},X_{(2)1}}&0\\
0&0&0&0\\
H_{X_{(1)2},M_{(2)1}}&0&H_{X_{(1)2},X_{(2)1}}&0
\end{array}
\right]
\end{equation}
where $H_{X_{(1)2},M_{(2)1}}$ represents the vertical M-X interactions, and has the form
\begin{equation}
H_{X_{(1)2},M_{(2)1}}=
\left[
\begin{array}{cccc}
X_s,M_s&0&0&X_s,M_{p_z}\\
0&0&0&0\\
0&0&0&0\\
X_{p_z},M_s&0&0&X_{p_z},M_{p_z}\\
\end{array}
\right].
\end{equation}

The elements themselves are
\begin{align}
X_s,M_s&=t^{(XM)}_{ss},\\
X_s,M_{p_z}&=t_{X_s-M_{p_z}}^{(XM)},\\
X_{p_z},M_s&=-t_{X_{p_z}-M_s}^{(XM)},\\
X_{p_x},M_{p_x}=&X_{p_y},M_{p_y}=t_{\pi}^{(XM)}\\
X_{p_z},M_{p_z}&=-t_{\sigma}^{(XM)}.
\end{align}

$H_{X_{(1)2},X_{(2)1}}$ represents the X-X interactions, with
\begin{equation}
H_{X_{(1)2},X_{(2)1}}=
\left[
\begin{array}{cccc}
X_s,X_s&X_s,X_{p_x}&X_s,X_{p_y}&X_s,X_{p_z}\\
X_{p_x},X_s&X_{p_x},X_{p_x}&X_{p_x},X_{p_y}&X_{p_x},X_{p_z}\\
X_{p_y},X_s&X_{p_y},X_{p_x}&X_{p_y},X_{p_y}&X_{p_y},X_{p_z}\\
X_{p_z},X_s&X_{p_z},X_{p_x}&X_{p_z},X_{p_y}&X_{p_z},X_{p_z}\\
\end{array}
\right].
\end{equation}

In the expressions for the elements, $L_c$ is the length of the inter-layer X-X hop, and is given by
\begin{equation}
L_c=\sqrt{\frac{a^2}{3}+\left(a_z-d_{XX}\right)^2}.
\end{equation}

The elements are
\begin{align}
\begin{split}
X_s,X_s&=f_1t_{ss}^{(XX)}\\
X_s,X_{p_x}&=-f_2\frac{a}{2L_c}t_{sp}^{(XX)}\\
X_s,X_{p_y}&=f_4\frac{a}{2\sqrt{3}L_c}t_{sp}^{(XX)}\\
X_s,X_{p_z}&=f_1\frac{a_z-d_{XX}}{2L_c}t_{sp}^{(XX)}\\
X_{p_x},X_s&=-X_s,X_{p_x}\\
X_{p_y},X_s&=-X_s,X_{p_y}\\
X_{p_z},X_s&=-X_s,X_{p_z}\\
X_{p_x},X_{p_x}&=f_1t_{\pi}^{(XX)}-f_3\left[\frac{a}{2L_c}\right]^2(t_{\pi}^{(XX)}+t_{\sigma}^{(XX)})\\
X_{p_y},X_{p_y}&=f_1t_{\pi}^{(XX)}-f_5\left[\frac{a}{2\sqrt{3}L_c}\right]^2(t_{\pi}^{(XX)}+t_{\sigma}^{(XX)})\\
X_{p_z},X_{p_z}&=f_1\left[t_{\pi}^{(XX)}-\left[\frac{a_z-d_{XX}}{L_c}\right]^2(t_{\pi}^{(XX)}+t_{\sigma}^{(XX)})\right]\\
X_{p_x},X_{p_y}&=-\frac{f_2}{\sqrt{3}}\left[\frac{a}{2L_c}\right]^2(t_{\pi}^{(XX)}+t_{\sigma}^{(XX)})\\
X_{p_x},X_{p_z}&=f_2\frac{a\left(a_z-d_{XX}\right)}{(2L_c)^2}(t_{\pi}^{(XX)}+t_{\sigma}^{(XX)})\\
X_{p_y},X_{p_z}&=-\frac{f_4}{\sqrt{3}}\frac{a\left(a_z-d_{XX}\right)}{(2L_c)^2}(t_{\pi}^{(XX)}+t_{\sigma}^{(XX)})\\
X_{p_y},X_{p_x}&=X_{p_x},X_{p_y}\\
X_{p_z},X_{p_x}&=X_{p_x},X_{p_z}\\
X_{p_z},X_{p_y}&=X_{p_y},X_{p_z}
\end{split}
\end{align}
In the case of the non-vertical M-X hops we have
\begin{equation}
H_{M_{(1)2},X_{(2)1}}=
\left[
\begin{array}{cccc}
M_s,X_s&M_s,X_{p_x}&M_s,X_{p_y}&M_s,X_{p_z}\\
M_{p_x},X_s&M_{p_x},X_{p_x}&M_{p_x},X_{p_y}&M_{p_x},X_{p_z}\\
M_{p_y},X_s&M_{p_y},X_{p_x}&M_{p_y},X_{p_y}&M_{p_y},X_{p_z}\\
M_{p_z},X_s&M_{p_z},X_{p_x}&M_{p_z},X_{p_y}&M_{p_z},X_{p_z}\\
\end{array}
\right]
\end{equation}
with matrix elements
\begin{align}
\begin{split}
M_s,X_s&=f_1^*t_{ss}^{(MX)}\\
M_s,X_{p_x}&=f_2^*\frac{a}{2L_M}t_{Ms-Xp}^{(MX)}\\
M_s,X_{p_y}&=-f_4^*\frac{a}{2\sqrt{3}L_M}t_{Ms-Xp}^{(MX)}\\
M_s,X_{p_z}&=f_1^*\frac{a_z-\frac{1}{2}(d_{XX}+d_{MM})}{2L_M}t_{Ms-Xp}^{(MX)}\\
M_{p_x},X_s&=-f_2^*\frac{a}{2L_M}t_{Mp-Xs}^{(MX)}\\
M_{p_y},X_s&=f_4^*\frac{a}{2\sqrt{3}L_M}t_{Mp-Xs}^{(MX)}\\
M_{p_z},X_s&=-f_1^*\frac{a_z-\frac{1}{2}(d_{XX}+d_{MM})}{2L_M}t_{Mp-Xs}^{(MX)}\\
M_{p_x},X_{p_x}&=f_1t_{\pi}^{(MX)}-f_3^*\left[\frac{a}{2L_M}\right]^2(t_{\pi}^{(MX)}+t_{\sigma}^{(MX)})\\
M_{p_y},X_{p_y}&=f_1t_{\pi}^{(MX)}-f_5^*\left[\frac{a}{2\sqrt{3}L_M}\right]^2(t_{\pi}^{(MX)}+t_{\sigma}^{(MX)})\\
M_{p_z},X_{p_z}&=f_1^*t_{\pi}^{(MX)}\\
&-f_1^*\frac{(a_z-\frac{1}{2}(d_{XX}+d_{MM}))^2}{L_M^2}(t_{\pi}^{(MX)}+t_{\sigma}^{(MX)})\\
M_{p_x},X_{p_y}&=-\frac{f_2^*}{\sqrt{3}}\left[\frac{a}{2L_M}\right]^2(t_{\pi}^{(MX)}+t_{\sigma}^{(MX)})\\
M_{p_x},X_{p_z}&=-f_2^*\frac{a\left(a_z-\frac{1}{2}(d_{XX}+d_{MM})\right)}{(2L_M)^2}(t_{\pi}^{(MX)}+t_{\sigma}^{(MX)})\\
M_{p_y},X_{p_z}&=\frac{f_4^*}{\sqrt{3}}\frac{a\left(a_z-\frac{1}{2}(d_{XX}+d_{MM})\right)}{(2L_M)^2}(t_{\pi}^{(MX)}+t_{\sigma}^{(MX)})\\
M_{p_y},X_{p_x}&=M_{p_x},X_{p_y}\\
M_{p_z},X_{p_x}&=M_{p_x},X_{p_z}\\
M_{p_z},X_{p_y}&=M_{p_y},X_{p_z}
\end{split}
\end{align}

\noindent where the length of the hop $L_M$ is given by
\begin{equation}
L_M=\sqrt{\frac{a^2}{3}+\left(a_z-\frac{1}{2}(d_{XX}+d_{MM})\right)^2}.
\end{equation}

For greater numbers of layers we build up the matrix such that $H_{1\mathrm{L}}$ is on the diagonal blocks, with adjacent diagonal blocks connected by $H_c$ and $H_c^{\dagger}$.

\section{Comparison between scissor corrected TB, uncorrected TB, and DFT optical matrix elements}
\label{SC_comp}
For comparison, we also obtain the matrix elements from density functional theory and from a TB model without scissor-correction. In a DFT calculation utilizing a plane-wave basis, the momentum matrix element is straightforward to calculate as
\begin{equation}
\mathbf{P}_{cv(v_1)}=\sum_{j} C^{\prime *}_{c,j} \cdot C^{\prime}_{v(v_1),j} \cdot \mathbf{G_j},
\end{equation}
\noindent where $C^{\prime}_{c,j},C^{\prime}_{v(v_1),j}$ and $G_{j}$ are the plane-wave coefficients and the reciprocal lattice vectors taken into account in the plane-wave basis set, respectively. $d_z$ is calculated in DFT by real space integration on a sufficiently fine grid. The intralayer TB parameters for the uncorrected model are given in Table \ref{parameters} in the main text, while the interlayer parameters are given in Table \ref{tab_multlayer_noSC}.
\begin{table}
\caption{Inter-layer hopping parameters (eV) for $\gamma$-InSe, without scissor-correction, as defined in the Hamiltonian (Eq. \eqref{bilayer_hamiltonian}).}
\label{tab_multlayer_noSC}
\begin{tabular}{ccccc}
\hline\hline
$t_{ss}^{(XX)}$&$t_{sp}^{(XX)}$& $t_{\pi}^{(XX)}$&$t_{\sigma}^{(XX)}$\\
$-0.731$&$-0.461$&$-0.119$&$-0.761$\\ \hline
$t_{ss}^{(MX)}$&$t_{X_s-M_{pz}}^{(MX)}$&$t_{X_{pz}-M_s}^{(MX)}$&$t_{\pi}^{(MX)}$&$t_{\sigma}^{(MX)}$\\
$-0.152$&$0.072$&$-0.504$&$0.198$&$0.015$\\ \hline 
$t_{ss}^{(XM)}$&$t_{X_s-M_{pz}}^{(XM)}$&$t_{X_{pz}-M_s}^{(XM)}$&$t_{\pi}^{(XM)}$&$t_{\sigma}^{(XM)}$\\
-0.332&$0.042$&$-0.208$&$-0.393$&$0.347$\\ \hline \hline
\end{tabular}
\end{table}

In the main text we apply a scissor correction to the DFT band structure in order to account for the underestimation of the band gap. It should be noted that the scissor correction is necessary for another reason as well. The TB model, when fitted to DFT band structures without scissor correction, agrees well with the DFT results for $d_z$ at small $N$ (see Fig. \ref{fig_dz_abs_TBSCvsDFT}a). However, the extrapolation to large $N$ is considerably larger, and the saturation much slower, than that for the TB model calculated with scissor correction. Likewise, the coefficient of the linear regime (Fig. \ref{fig_alpha_TBSCvsDFT}) is also larger if the scissor correction is omitted. In essence, uncorrected DFT overestimates $d_z$ matrix elements, whereas the scissor correction leads to a significant reduction in $d_z$, and in turn a reduced absorption.

The origin of the reduction in $d_z$ when applying the scissor correction lies in the effect the scissor correction has on the TB wave functions for $N>1$. Fig. \ref{fig_TBWF_N19}a shows the modulus square of the chalcogen $p_z$ TB wave function coefficients as a function of the sublayer index in 19-layer InSe. The wave function reduces towards the edge of the slab but a clear finite value remains at the very edge, due to a substantial oscillation in the coefficients, which gives a substantial contribution to $d_z$. Fig. \ref{fig_TBWF_N19}b shows the same wave function coefficients after scissor correction. Note that the relative weight of the coefficient at the edge has now decreased, as has the aforementioned oscillation, which leads to an overall smaller $d_z$ and a faster saturation with increasing $N$.

The physics behind the reduction of the wave function at the edge is the relative reduction of the inter-layer interaction as compared to the intra-layer interaction, which is a direct consequence of the scissor correction. By increasing the gap without changing the band width caused by inter-layer hopping, the intra-layer hopping becomes stronger while the inter-layer hopping remains at the same magnitude. This can be understood within a chain model, which we discuss below.

The core message here is that the scissor correction has an effect on the wave functions and in turn on the optical properties of InSe slabs, and should be taken into account when modeling few-layer InSe.

\begin{figure}
 \centering
   \includegraphics[width=0.45\textwidth]{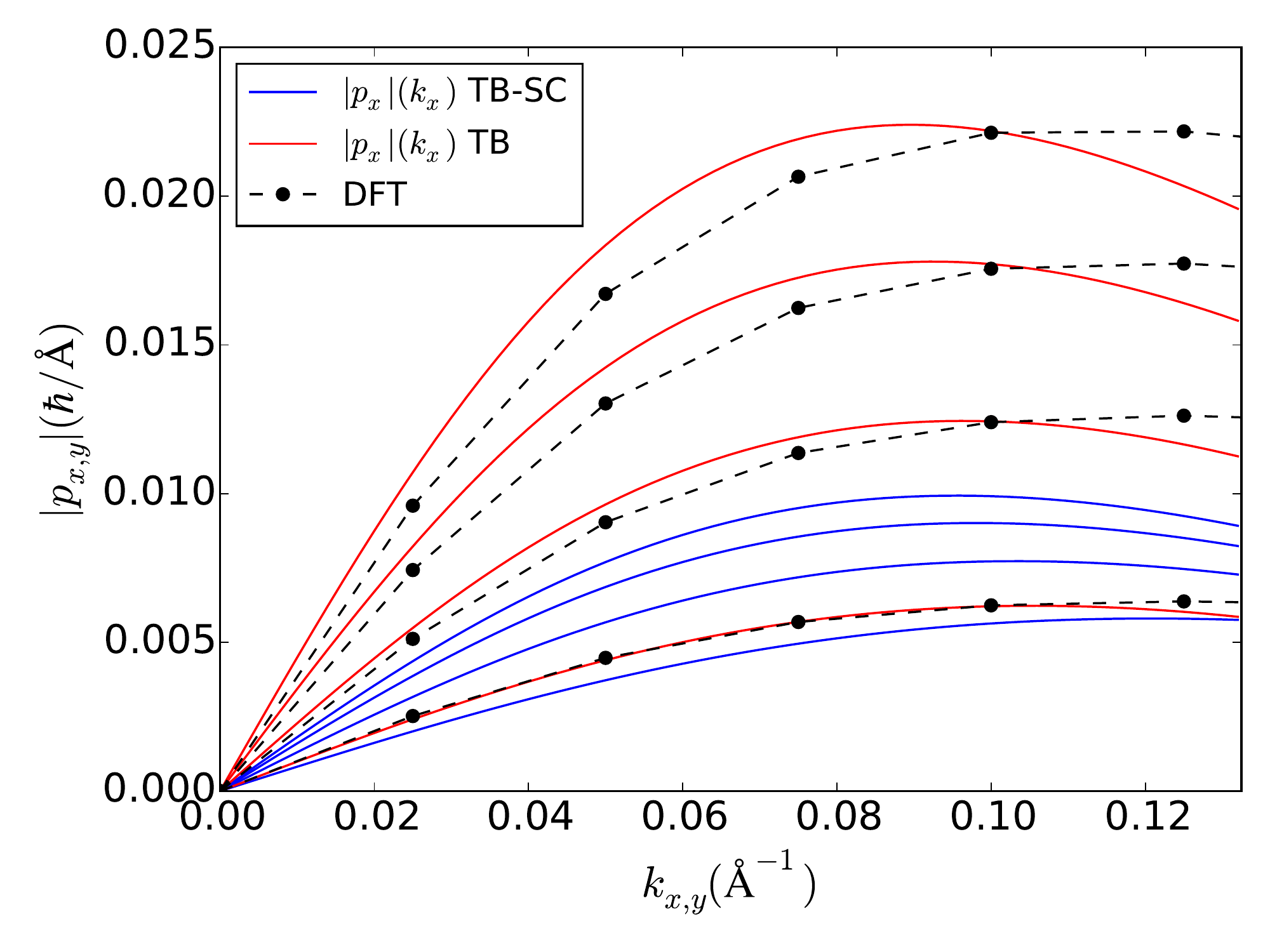}
\caption{(Color online) Comparison between the TB and DFT matrix elements for the $x$ components of the interband momentum for  $N=$ 2, 3, 4 and 5 layer $\gamma$-InSe.}
\label{fig_p_matr_el_TBSCvsDFT}
\end{figure}

\begin{figure}
 \centering
   \includegraphics[width=0.45\textwidth]{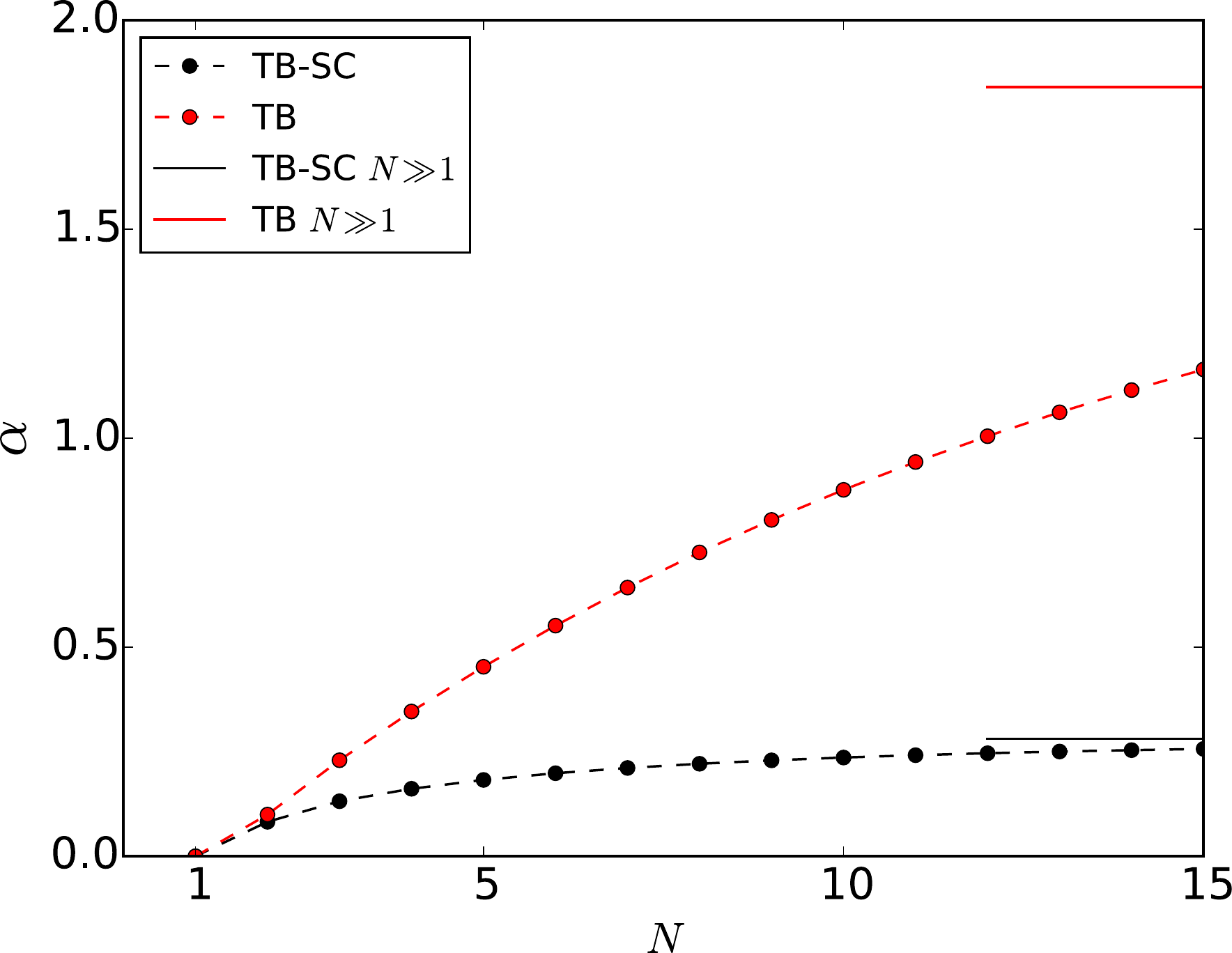}
\caption{(Color online) Comparison between the scissor corrected and uncorrected TB results for the $\alpha$ parameter.}
\label{fig_alpha_TBSCvsDFT}
\end{figure}

\begin{figure}
 \centering
   \includegraphics[width=0.45\textwidth]{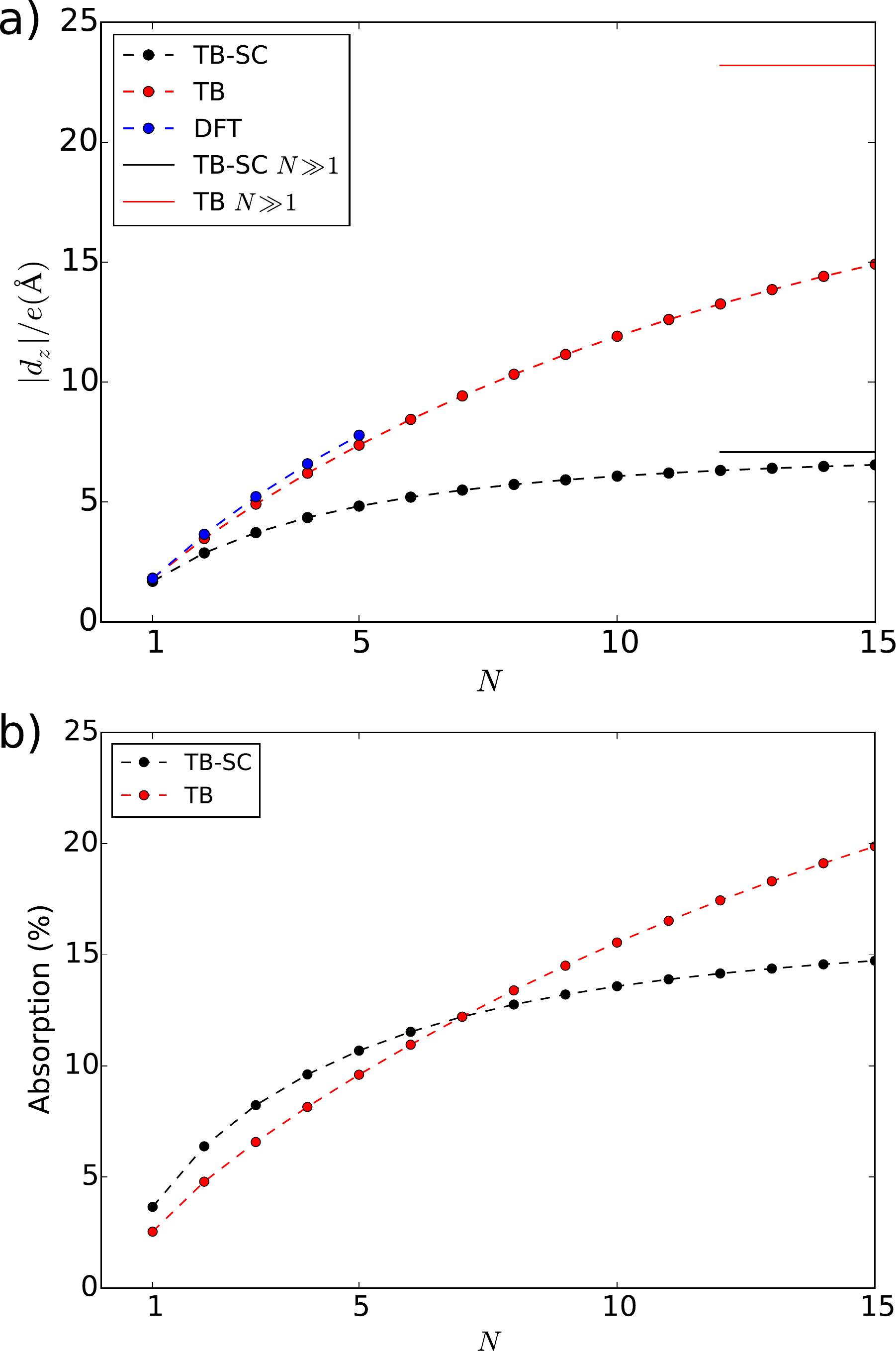}
\caption{(Color online) Comparison between the scissor corrected and uncorrected TB results for the $d_z$ matrix element (a) and the band edge absorption (b).}
\label{fig_dz_abs_TBSCvsDFT}
\end{figure}

\begin{figure}
 \centering
   \includegraphics[width=0.45\textwidth]{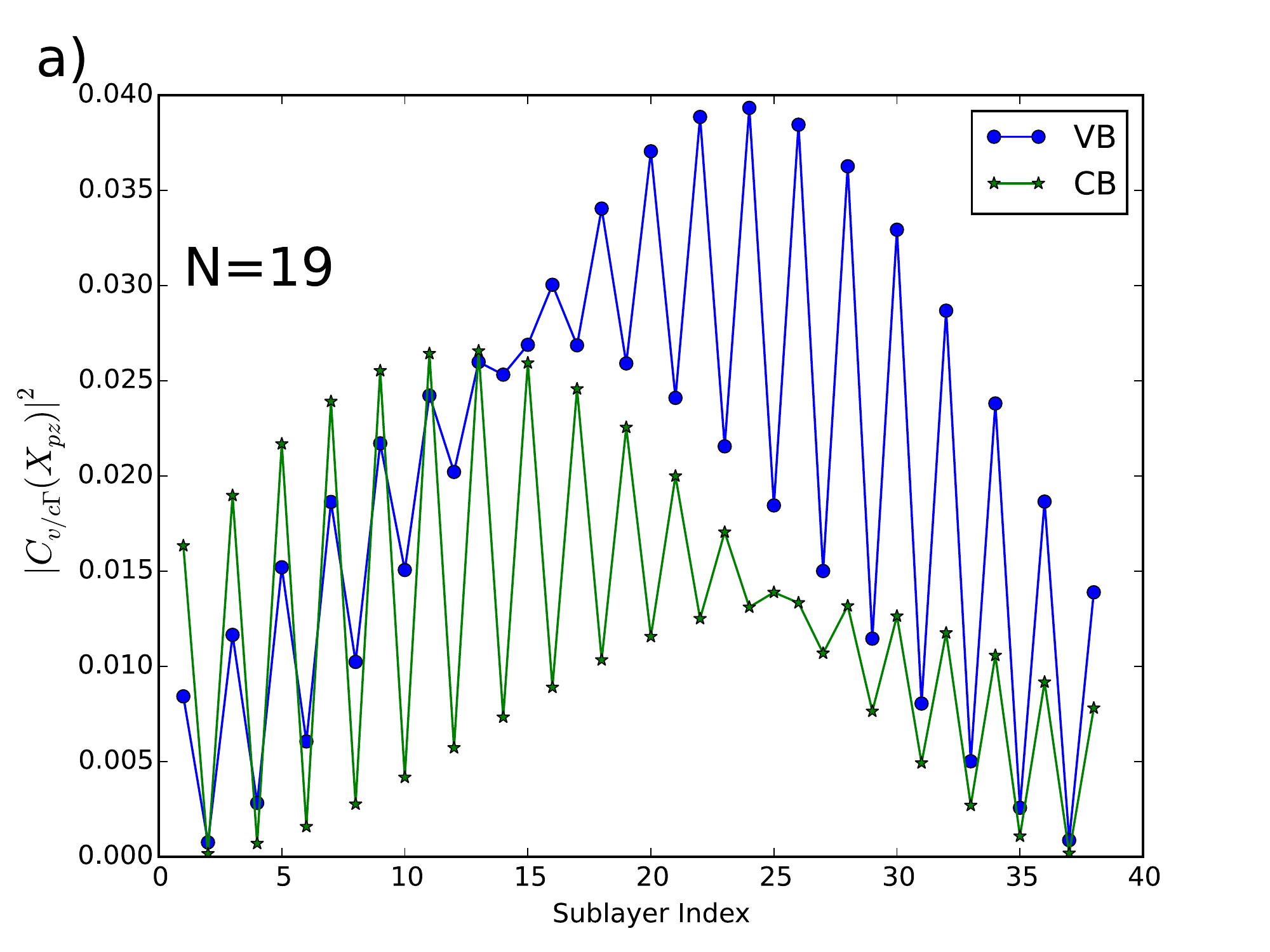}\\
   \includegraphics[width=0.45\textwidth]{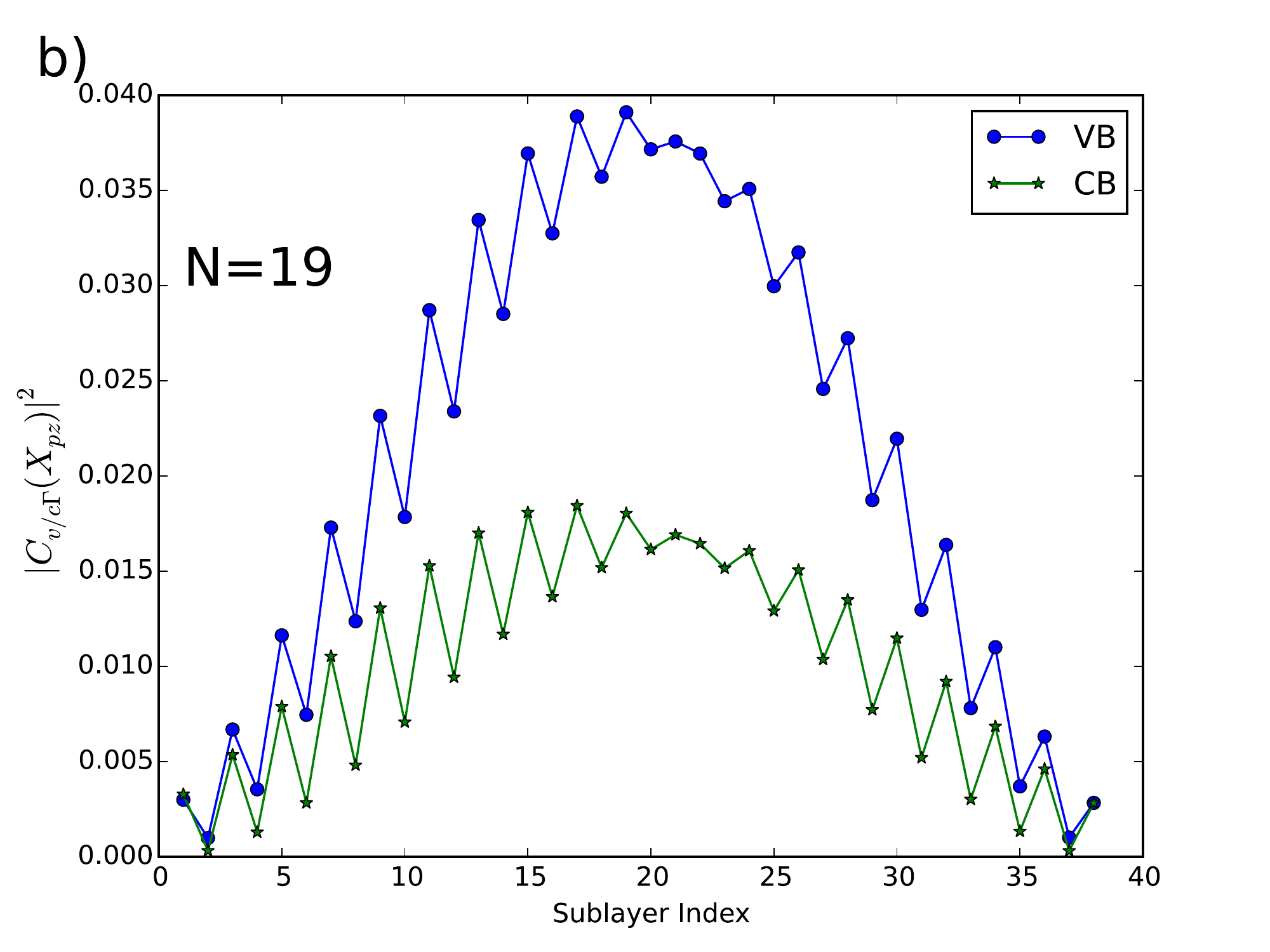}
\caption{(Color online) Distribution of the $X_{pz}$ TB wave function coefficients along the slab in 19-layer InSe, without (a) and with (b) scissor correction.}
\label{fig_TBWF_N19}
\end{figure}

\section{Chain model for few-layer InSe at $\Gamma$}

The simplest way to describe a layered semiconductor is by approximating each layer with a dimer, each atom hosting a single basis orbital $x_1$ and $x_2$. In this case, the monolayer can be described by a single hopping integral $t$ and the Hamiltonian will be

\begin{equation}
H=t\sum_{i}x_{2i}^{\dagger}x_{1i}+h.c.
\end{equation}

\noindent where  $x_{2i}^{(\dagger)}$ annihilates (creates) an electron on sublayer 2, site $i$. Expressed in matrix form the Hamiltonian is

\begin{equation}
H=\left[
\begin{array}{cc}
0&t\\
t&0\\
\end{array}
\right].
\end{equation}

The Hamiltonian of a few-layer structure is

\begin{align}
\begin{split}
H=\sum_{i}&\left[t\sum_{(n)}(x_{(n)2i}^{\dagger}x_{(n)1i})\right.\\
&\left.+t^{\prime}\sum_{(n)=1,2}^{N-1}(x_{(n)2i}^{\dagger}x_{(n+1)1i})\right]+h.c.
\end{split}
\end{align}

\noindent where $(n)$ is the layer index, and the inter-layer interaction is described by the hop $t^{\prime}$. As an example, the matrix form of a bilayer can be written as

\begin{equation}
H=\left[
\begin{array}{cccc}
0&t&0&0\\
t&0&t'&0\\
0&t'&0&t\\
0&0&t&0\\
\end{array}
\right],
\end{equation}

\noindent which has the following eigenvalues:

\begin{align}
\begin{split}
&\frac{1}{2}\left(t'+\sqrt{4t^2+t'^2}\right),\\
&\frac{1}{2}\left(-t'+\sqrt{4t^2+t'^2}\right),\\
&\frac{1}{2}\left(t'-\sqrt{4t^2+t'^2}\right),\\
&\frac{1}{2}\left(-t'-\sqrt{4t^2+t'^2}\right).
\end{split}
\end{align}

In this chain model, the ratio $t^{\prime}/t$ characterizes the strength of the inter-layer interaction with respect to the intra-layer coupling. Let us now assume that we can describe few-layer InSe with such a model, with some values for the two hopping parameters obtained from DFT calculations. When we implement a scissor correction, we leave the inter-layer hop $t^{\prime}$ unchanged while we increase the magnitude of $t$ since, in the monolayer, the band gap from this model is simply $2t$. Hence, a scissor correction translates to a decrease in the ratio $t^{\prime}/t$.

\begin{figure}
 \centering
   \includegraphics[width=0.45\textwidth]{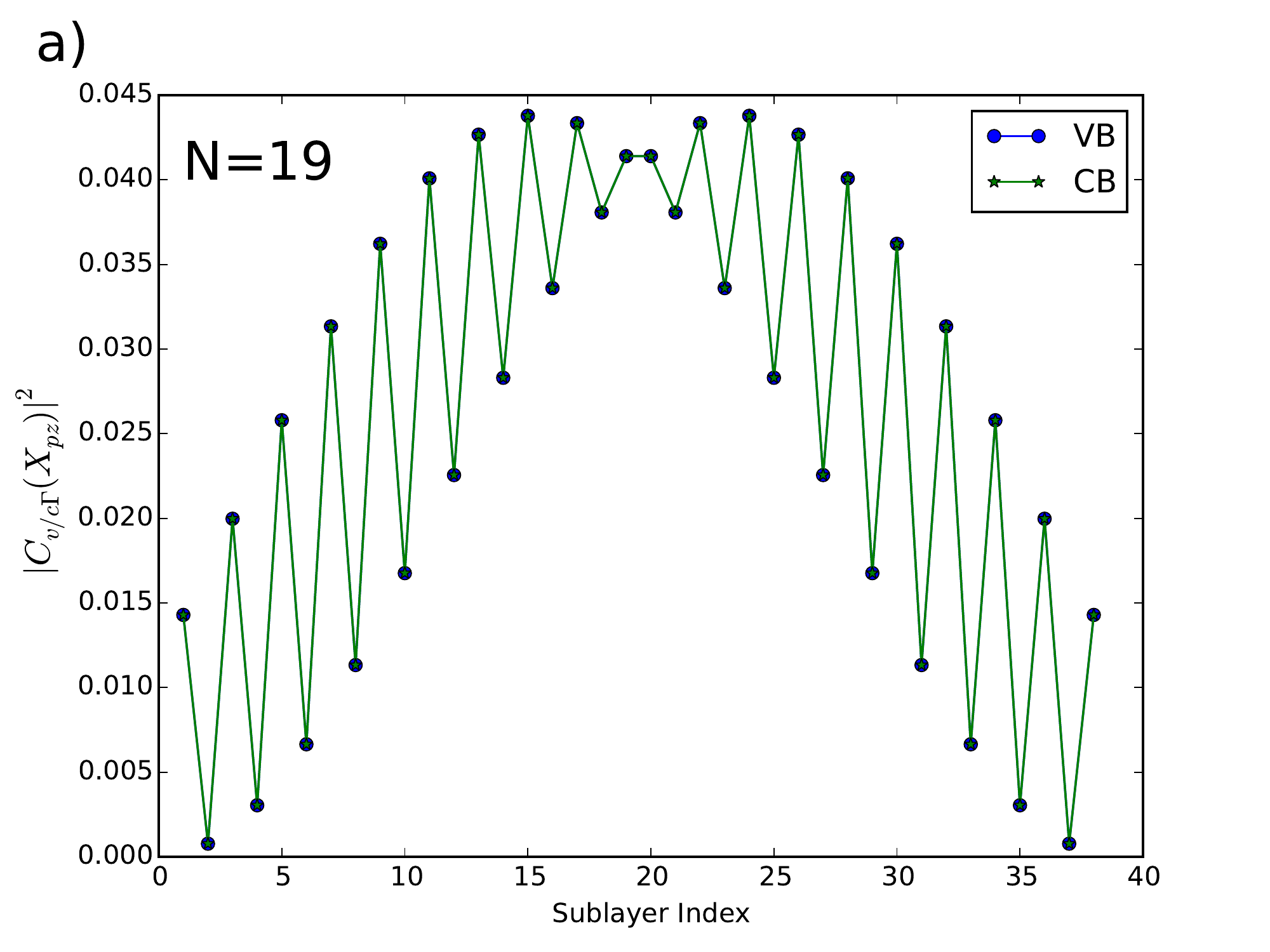}\\
   \includegraphics[width=0.45\textwidth]{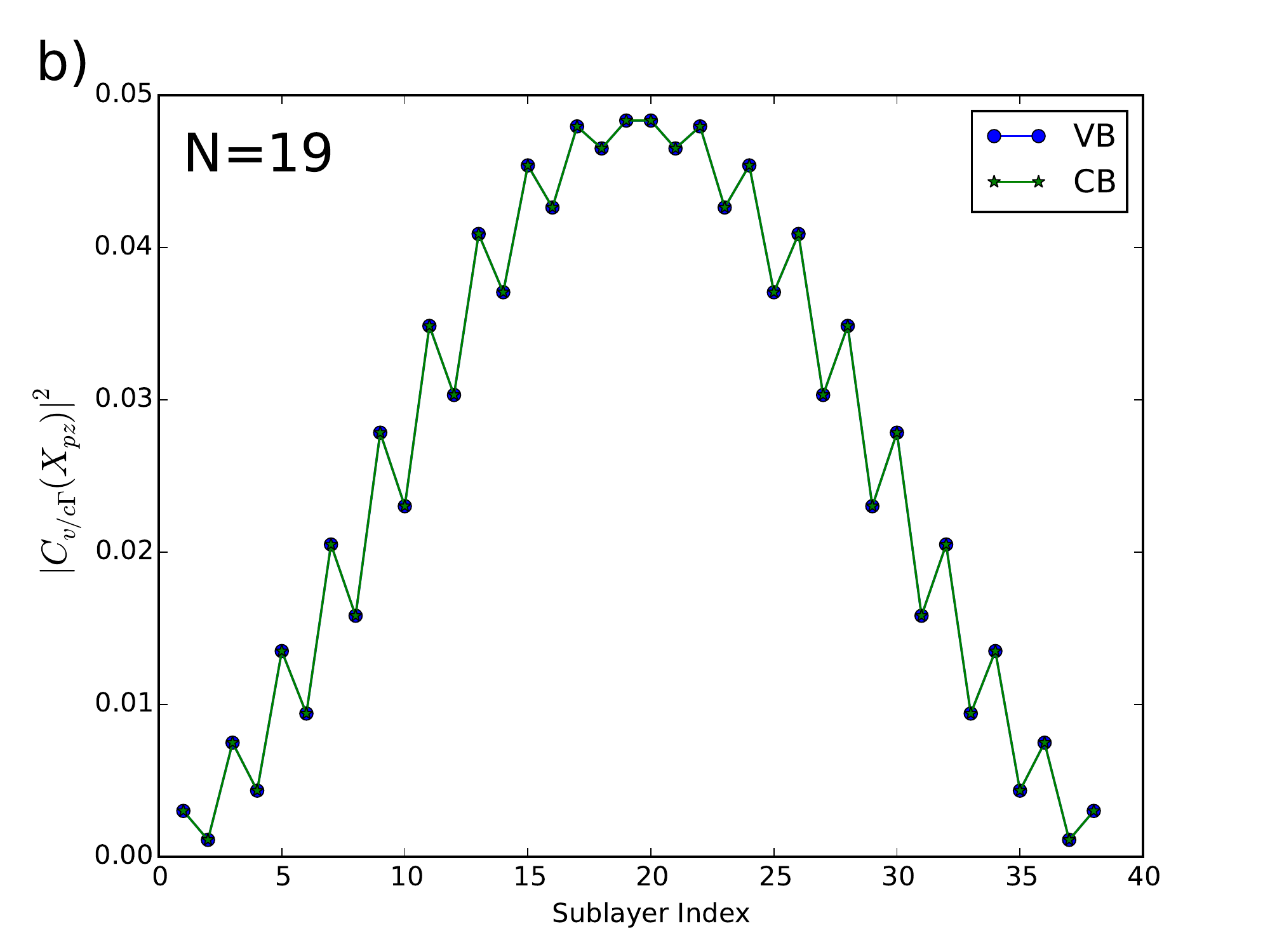}
\caption{(Color online) Distribution of the wave function coefficients along the slab in 19-layer InSe according to the chain model, with hopping ratios $t^{\prime}/t=0.8$ (a) and $t^{\prime}/t=0.4$ (b).
}
\label{fig_TBWF_N19_tm}
\end{figure}

\begin{figure}
 \centering
   \includegraphics[width=0.45\textwidth]{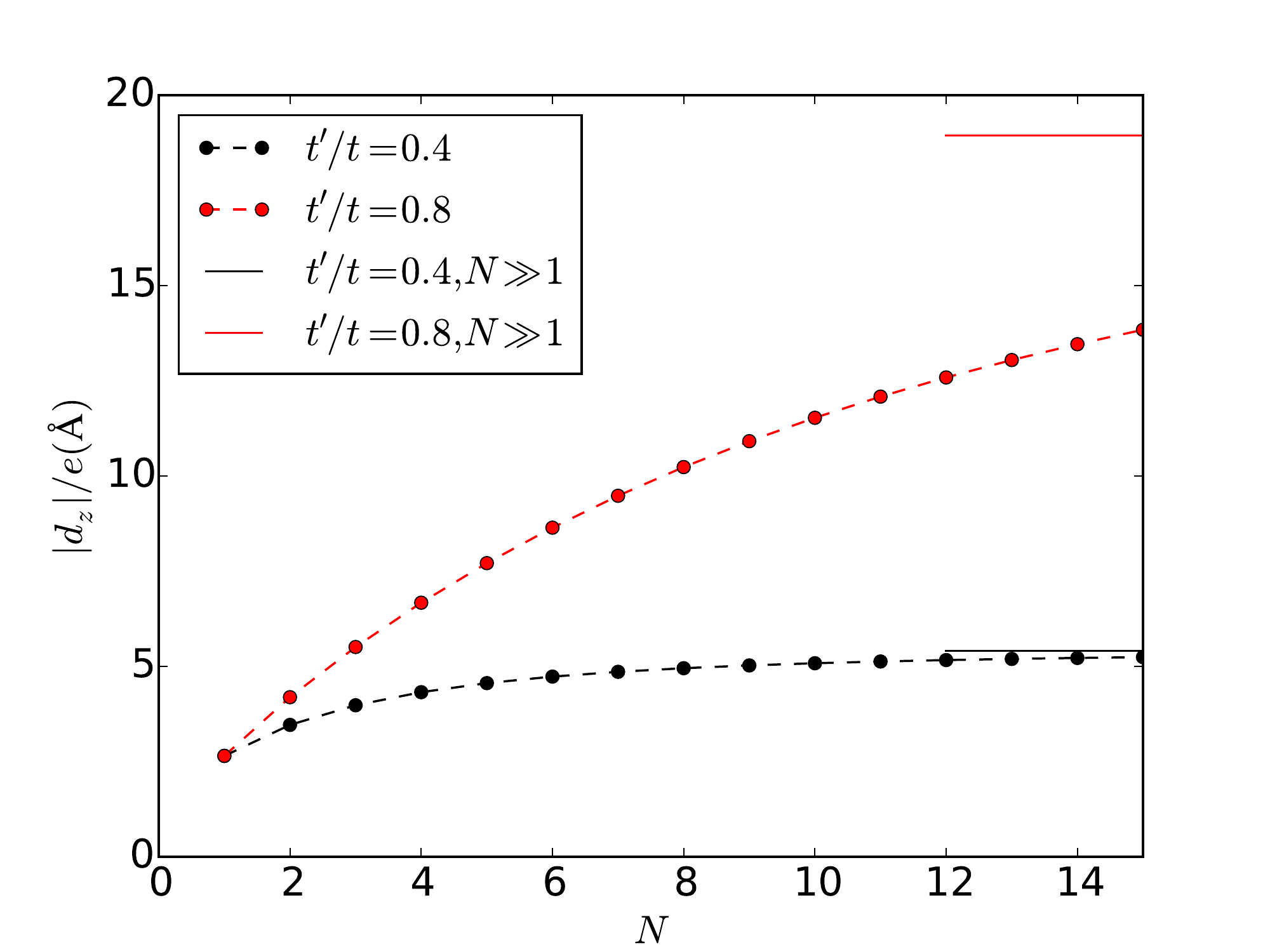}
\caption{(Color online) The $d_z$ matrix element of few-layer InSe according to the chain model, with hopping ratios $t^{\prime}/t=0.8$ and $t^{\prime}/t=0.4$.}
\label{fig_dz_tm}
\end{figure}

This finding allows us to demonstrate the qualitative effect of the scissor correction. Fig. \ref{fig_TBWF_N19_tm} shows the modulus square of the coefficients $C_{v/c}$ of the chain model wave functions in the valence and conduction band. Panel a) corresponds to $t^{\prime}/t=0.8$, while panel b) to $t^{\prime}/t=0.4$. The visible reduction of the wave function along the edges upon decreasing $t^{\prime}/t$ is in agreement with the effects of the scissor correction on the full model  (see Fig. \ref{fig_TBWF_N19}). Similarly, if we now plot the $d_z$ matrix element from the chain model (Fig. \ref{fig_dz_tm}) we find that the matrix element undergoes significant reduction when we decrease $t^{\prime}/t$, just like it happened in the full model when we implemented the scissor correction there (see Fig. \ref{fig_dz_abs_TBSCvsDFT}a).

%

\end{document}